\documentclass[sigconf]{acmart}

\usepackage{color}
\usepackage{rotating}
\usepackage{tabularray}
\usepackage{subcaption}
\usepackage{dirtytalk}
\usepackage{xcolor}
\usepackage{mdframed}
\usepackage{enumitem}

\AtBeginDocument{%
  }

\setcopyright{acmlicensed}
\copyrightyear{2026}
\acmYear{2026}
\acmDOI{XXXXXXX.XXXXXXX}
\acmConference[UIST '26]{ACM (Association of Computing Machinery) Symposium on User Interface Software and Technology}{November 2026}{Detroit, MI}
\acmISBN{978-1-4503-XXXX-X/2018/06}

\begin{document}

\title{Exploring the Interaction of Explanation Styles, Context, and Trust of AI Privacy Redaction in AI-mediated Interactions}

\author{Roshni Kaushik}
\affiliation{%
  \institution{Fujitsu Research of America}
  \city{Pittsburgh}
  \state{Pennsylvania}
  \country{USA}
}
\email{rkaushik@fujitsu.com}
\author{Maarten Sap}
\affiliation{%
  \institution{Carnegie Mellon University}
  \city{Pittsburgh}
  \state{Pennsylvania}
  \country{USA}
}
\email{msap2@andrew.cmu.edu}
\author{Koichi Onoue}
\affiliation{%
  \institution{Fujitsu Research of America}
  \city{Pittsburgh}
  \state{Pennsylvania}
  \country{USA}
}
\email{konoue@fujitsu.com}

\renewcommand{\shortauthors}{Kaushik, et. al.}

\begin{abstract}
  Humans are increasingly using AI-mediated communication to help facilitate human interactions; however, in privacy-sensitive domains, the AI mediator has the additional challenge of considering how to maintain user trust and adapt to user preferences while preserving privacy.
Prior work has shown explanations of redaction improve trust compared to no explanations in privacy-sensitive scenarios. We explore this research area further by developing a system where an AI mediates conversations containing private information while providing explanations of different styles. The system uses an LLM for (1) information generation in a particular domain and with a specific amount of sensitive information to redact, (2) redaction of sensitive information, (3) explanation generation in a particular style, and (4) redaction of sensitive information in the generated explanation. We then conduct an online user study ($n=249$) to understand the interaction between explanation styles, context (domain and redaction amount), and trust in these privacy-sensitive scenarios, where users choose which explanation style(s) they prefer for a particular context and rate their trust in the system with their preferred explanation(s). 
Our results reveal that explanation preferences vary systematically with context, users' trust in the system is linked to preferred explanation, and users display consistent individual differences in how they choose explanations and how their preferences depend on context. 
These findings highlight the importance and design implications of developing personalized, context-aware explanations in privacy-sensitive scenarios and suggest that adaptive explanations are essential for designing privacy-aware and trustworthy AI mediators.

\end{abstract}



\received{13 January 2026}

\maketitle

\definecolor{ResearcherColor}{HTML}{A02B93}
\definecolor{CoordinatorColor}{HTML}{156082}
\definecolor{CollaboratorColor}{HTML}{196B24}

\section{Introduction}
\label{sec:intro}
AI systems are increasingly used to mediate communication between people, particularly in settings where sensitive information must be selectively revealed or withheld (e.g., healthcare, finance, and social media).
For example, AI systems are already used to simplify texts for broader audiences \cite{araujo_simplifying_2023} and to remove sensitive information from healthcare records \cite{caine_patients_2013}. In these contexts, an AI mediator must modify or filter content to preserve privacy while enabling meaningful interaction between humans. 
It actively alters the messages exchanged between users, and as a result, these systems are also tasked with maintaining trust and shared understanding between the humans involved. 
Explanations can help clarify why certain information was withheld and help users interpret the system's behavior, supporting users in calibrating their trust and understanding of AI outputs \cite{kim__2023, haque_explainable_2025}. However, explanation interfaces are often designed with a fixed format or a single explanation type, implicitly assuming that one explanation approach will be effective for all users across different situations and domains.

User preferences for how AI mediators construct and present explanations can vary greatly depending on context, task, and the nature of the information being conveyed. Prior work suggests that different explanation styles can lead to different levels of understanding, satisfaction, and perceived usefulness \cite{ribeiro__2016, kulesza_too_2013}, and that users may respond differently to system behavior when information is incomplete or modified, such as in cases of filtering or refusal \cite{zheng_let_2025, wester_as_2024}. In AI-mediated settings, variations in domain or the extent of information redaction may further shape these preferences, as users must interpret outputs that are altered by privacy requirements. Additionally, explanation choice may directly influence users' trust in the system, affecting how they interpret AI-mediated content \cite{hoffman_metrics_2019}. However, there remains limited systematic understanding of how context, trust, and individual differences jointly shape explanation preferences, particularly in privacy-sensitive scenarios where information is intentionally withheld.

To address this gap, we investigate how to design AI-mediated interactions that support user preferences over explanations in contexts with varying information constraints. We focus on scenarios where an AI mediator selectively redacts or withholds private information that the user should not have access to. We conduct a user study in which online participants ($n=249$) interact with an AI mediator across multiple domains and levels of redaction (e.g., a large amount of sensitive information requiring heavy redaction vs. little sensitive information requiring light redaction), and are given the ability to select among different explanation types. This setup allows us to examine how explanation preferences vary with context, and how these choices relate to trust, individual differences, and user feedback. By jointly considering contextual factors (domain and redaction level) and user characteristics, our approach aims to inform the design of adaptive explanation interfaces that better support transparency, trust, and effective human-AI interaction.

Building on this goal, we investigated five key research questions addressing explanation choice, individual differences, trust, and feedback. We first examine how \textbf{context influences explanation choice} (RQ1) and how these choices affect \textbf{user trust} (RQ2). We then explore \textbf{individual differences} in explanation selection (RQ3) and how \textbf{user characteristics correlate with explanation preferences and trust} (RQ4). Finally, we consider how users want to \textbf{provide feedback on explanations} to improve system behavior (RQ5).

Our study reveals several key insights about how users interact with AI mediators and translates these findings into actionable contributions. We find that \textbf{explanation preferences vary systematically with context}, indicating that one-size-fits-all explanations are often insufficient. Users' \textbf{trust in the system is closely linked to using their preferred explanations}, highlighting the importance of giving users their preferred explanation styles to foster trust. Beyond contextual effects, we identify \textbf{distinct user-specific patterns in explanation selection}, suggesting that explanation preferences are not only situational but also shaped by individual characteristics. Finally, participants expressed clear preferences for how they would like to \textbf{provide feedback on explanations}, offering guidance for designing interactive and responsive AI systems.

This work makes the following three primary contributions.
\begin{enumerate}
    \item Design implications for AI-mediated interactions, including strategies for supporting user preferences for explanations, adapting explanations to context, and collecting user feedback effectively
    \item Empirical insights into how explanation preferences and trust vary across domains and levels of information redaction
    \item Characterization of individual differences in explanation preferences and trust, providing guidance for personalized explanation design
\end{enumerate}

 These three contributions inform the design of AI-mediated interfaces that support appropriate user trust, maintain transparency, and enable effective human-AI interaction in privacy-sensitive scenarios.

\section{Related Work}
\label{sec:related_work}
\subsection{Explainable AI and Human Evaluation of Explanations}

Explainable AI (XAI) aims to make model behavior understandable to users and supports transparency and users' decision making \cite{adadi_peeking_2018, arrieta_explainable_2019}. Researchers have developed a wide range of explanation techniques, including feature-based approaches \cite{ribeiro__2016}, example-based explanations \cite{cai_effects_2019}, and counterfactual explanations that describe how outcomes could change under different inputs \cite{verma_counterfactual_2024}. Explanations can improve how users interpret and understand the system's outputs, and prior work has explored the effects of different styles of explanations \cite{miller_explanation_2019}.

However, explanations are also interactive elements that shape how users reason about systems. HCI research emphasizes evaluating explanations based on their ability to support user decision making, understanding, and interaction \cite{abdul_trends_2018}. Prior work highlights the importance of tailoring explanations to user goals and context \cite{lim_why_2009}, and later work shows that well-designed explanations can help users form a more accurate understanding of the system and improve task performance \cite{kulesza_too_2013, kulesza_principles_2015}. These findings suggest that explanations should be treated as part of the user experience and should be adaptable rather than static.

Prior work has proposed a variety of explanation quality evaluation methods, including subjective measures such as perceived usefulness and satisfaction, as well as objective measures such as task performance and mental model accuracy \cite{ribeiro__2016, kulesza_too_2013}. Other researchers further distinguish between different evaluation approaches \cite{doshi-velez_towards_2017}, which highlights the importance of studying explanations in realistic usage contexts. However, many studies continue to evaluate explanations in isolation, often presenting a single explanation type without considering how different users might prefer different styles in different situations.

A key limitation of this body of work is its assumption of one-size-fits-all explanations. In practice, different explanation types may be more or less effective depending on the task, domain, and user characteristics \cite{ribeiro__2016}. Relatively little is known about how users could choose between multiple explanation options, particularly in contexts where information is incomplete or redacted. Addressing this gap requires moving beyond evaluating individual explanations towards understanding explanation generation as an interactive, context-dependent process.

\subsection{Explanations and Trust in AI Systems}

Trust is incredibly important in human-AI interaction, influencing whether users rely on, ignore, or appropriately calibrate their use of intelligent systems. Prior work has shown that transparency and explanations can play a key role in shaping user trust, helping users understand system behavior \cite{lee_trust_1992, lee_trust_2004}. In these works, explanations are usually positioned as a mechanism for calibrating trust, and they can encourage appropriate reliance instead of over or under trust \cite{jacovi_formalizing_2021}.

Researchers have also examined how different types of explanations affect user trust and reliance on systems. Some studies find that explanations can increase perceived transparency and user confidence \cite{ribeiro__2016, kizilcec_how_2016}. However, other studies have found more nuanced effects; for example, explanations may lead to overtrust if they create an illusion of understanding, or undertrust when users find them unhelpful or difficult to interpret \cite{poursabzi-sangdeh_manipulating_2021, kim__2023, park_critical_2025}. These mixed findings suggest that the relationship between explanations and trust is not straightforward, and it depends on factors such as the task, explanation style, and user expectations.

Recent work has also emphasized trust calibration, where users align their level of trust with the actual capabilities and limitations of the system \cite{jacovi_formalizing_2021}. In this perspective, effective explanations are those that increase trust, help users develop appropriate mental models, and allow users to make informed decisions about when to rely on AI systems. This viewpoint shifts the focus from only maximizing trust to promoting more informed and context-aware interactions.

Much of existing literature treats explanations as a fixed intervention and evaluates their impact on trust in a single interaction. There has been less work looking at how user preferences for different explanations relate to trust, or how trust evolves when users are given explanations that align with their needs and preferences. Understanding this relationship is especially important in scenarios where AI systems mediate and/or redact information, as users are utilizing the explanations to understand the system's interventions and what information might be missing or changed.

\subsection{Adaptive, Personalized, and User-Driven Explanation Interfaces}

Prior work has explored adaptive and personalized approaches to tailor explanations to different users and contexts. Personalization often utilizes user characteristics such as domain expertise, prior knowledge, or cognitive preferences to provide more relevant and effective explanations \cite{lage_human_2019}. For example, some expert users might prefer detailed, feature-based explanations, while novice users might benefit from simpler or example-based explanations. These approaches highlight the importance of aligning explanation design with individual user needs, especially since misalignment could lead to refusals \cite{wester_as_2024, zheng_let_2025}.

Researchers have independently also investigated context-aware explanation systems that adapt explanations based on situational factors. Work in intelligibility suggests that users require different types of explanations depending on their goals, such as understanding system behavior or justifying outcomes \cite{lim_why_2009}. Other systems dynamically adjust explanation content or format based on the context of the interaction, aiming to improve usability \cite{zein_personalized_2025, zhang_towards_2025}. This work emphasizes that effective explanations depend on who the user is, the context of the interaction, and what the users are trying to accomplish.

Another line of research explores user-driven explanation interfaces which give users control over how explanations are presented. These systems can allow users to select, switch between, or request different types of explanations on demand \cite{kulesza_principles_2015, ribeiro__2016}. Providing such control can support exploration, help users form more accurate mental models, and accommodate diverse preferences \cite{cappuccio_explanation_2025}. However, this approach introduces new challenges, including how users make choices among different explanation options and the explanations they select to best support their understanding.

Prior work has explored both adaptive and user-driven approaches, but these directions are often explored separately. Adaptive systems change their approach based on context, and user-driven interfaces adapt to individual preferences, but the interaction between these two is rarely explored. As a result, relatively little is known about how user characteristics, contextual factors, and explanation styles interact in practice, especially in scenarios where the full information is not presented to users.

\subsection{Feedback and AI-Mediated Interaction under Information Constraints}

Many real-world AI systems operate as mediators that filter, summarize, or redact information to satisfy constraints such as privacy, confidentiality, or safety. Prior work on privacy-preserving systems highlights the need to balance utility with the protection of sensitive data \cite{nissenbaum_privacy_2004}. Techniques such as differential privacy and automated redaction attempt to limit the exposure of sensitive information, but can also introduce uncertainty by altering or omitting relevant details. As a result, users interacting with these types of systems must reason about what is shown and what has been changed or removed by the system.

These information constraints introduce new challenges for transparency and user understanding. Research on privacy-aware interfaces suggests that users often struggle to interpret system behavior when data is selectively withheld, which can lead to reduced trust, incorrect assumptions, and confusion \cite{kocielnik_will_2019}. There have been calls for research on how LLM design paradigms affect disclosure behaviors, mental models and preferences for privacy control, and the design of tools that empower users to reclaim ownership over their personal data \cite{li_human-centered_2024}. Providing appropriate explanations in these settings is therefore critical, but is also difficult, as explanations may be constrained by the same privacy requirements.

Additionally, prior work has explored mechanisms for incorporating user feedback into AI systems, particularly in interactive and human-in-the-loop settings. Feedback can be explicit, such as ratings or corrections, or implicit, such as behavioral signals, and is often used to refine system outputs or improve personalization \cite{amershi_power_2014}. In the context of explainable AI, researchers have proposed feedback mechanisms to help users refine explanations, which enables iterative improvement of the model behavior and explanation quality \cite{kulesza_principles_2015}. However, most of this work assumes users have full visibility of the system's outputs and does not account for scenarios where information can be redacted.

In summary, prior research highlights the importance of explanations, trust, personalization, and feedback in human-AI interaction, as well as the challenges introduced by privacy considerations. However, the interaction between these research areas has been under-explored. Specifically, little is known about how users select explanations, how those choices relate to trust and individual differences, and how context is relevant in these interactions, especially when the system's role is to redact private information. Addressing these gaps requires a more integrated, human-centered understanding of explanations in realistic settings with AI-mediation. In this work, we take a step in that direction by examining how context, user characteristics, and different explanation styles interact to influence trust and preferences for privacy-sensitive scenarios.

\section{AI-Redaction System Design}
\label{sec:system_design}
\begin{figure*}[ht]
    \centering
    \includegraphics[width=2\columnwidth]{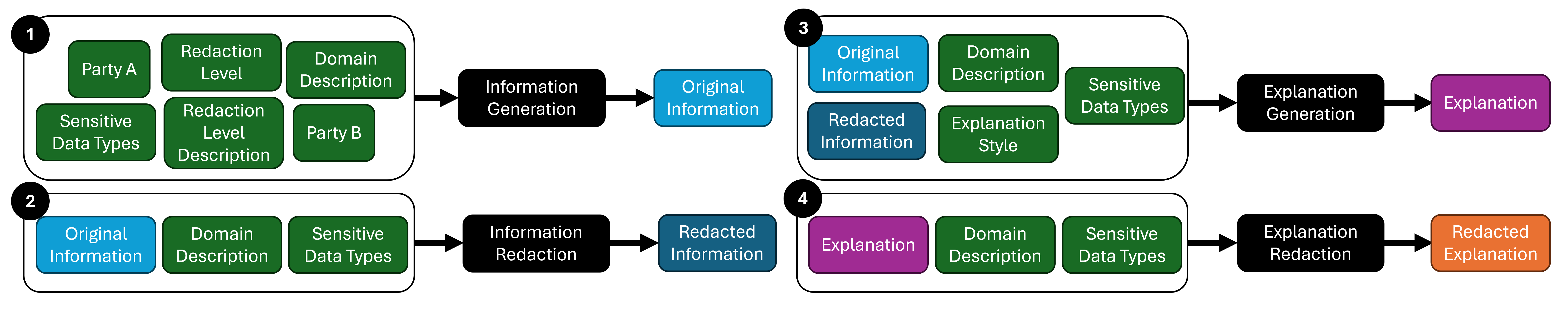}
    \caption{System Diagram showing the steps for (1) Information Generation, (2) Information Redaction, (3) Explanation Generation, and (4) Explanation Redaction}
    \label{fig:system_diagram}
\end{figure*}

Our goal is to determine how user explanation preferences change based on different domains and redaction levels as well as user trust for each scenario. We used GPT-5 for all generation steps, with a single prompt for each of the four steps depicted in Figure \ref{fig:system_diagram}. We used Gemma3 for any LLM judges used for evaluation to prevent any bias introduced by using the same model for generation and evaluation.

\subsection{Step 1: Information Generation}
The information generation step takes as input a set of variables defining the domain. We explored $6$ domains, as defined in Table \ref{tab:domain_information} in Appendix \ref{sec:appendix-domain_information}, which are medical, education, workplace, social media, financial, and legal. References justifying the choice of these domains are cited in the table, and these domains are also common areas where users could have privacy preferences. Some domains, such as medical and financial, have the potential for greater privacy sensitivity due to criticality of information, while others could be viewed as less critical, such as social media. For each domain, we define the individual with access to private information (Party A) and the individual they are communicating with who does not have access (Party B). Next we detail a description that explains the interaction between the two parties in this domain. Lastly, we create a list of $7$ sensitive data types for each domain that could be included in any generated information. These information types should not be accessed by Party B. 

We also have to define the amount of sensitive information that will be present in the generated information. We define three redaction levels: heavy, moderate, and light, which contain little, some, and more sensitive information, motivated by prior work which indicates that explanation preferences and trust can vary based on the redaction level \cite{kaushik2026examiningeffectexplanationsai}. Specifically, we define the redaction levels for this work as follows:
\begin{itemize}
    \item Heavy: contains 5-7 types of sensitive data
    \item Moderate: contains 3-5 types of sensitive data
    \item Light: contains 1-3 types of sensitive data
\end{itemize}

\begin{figure*}[ht]
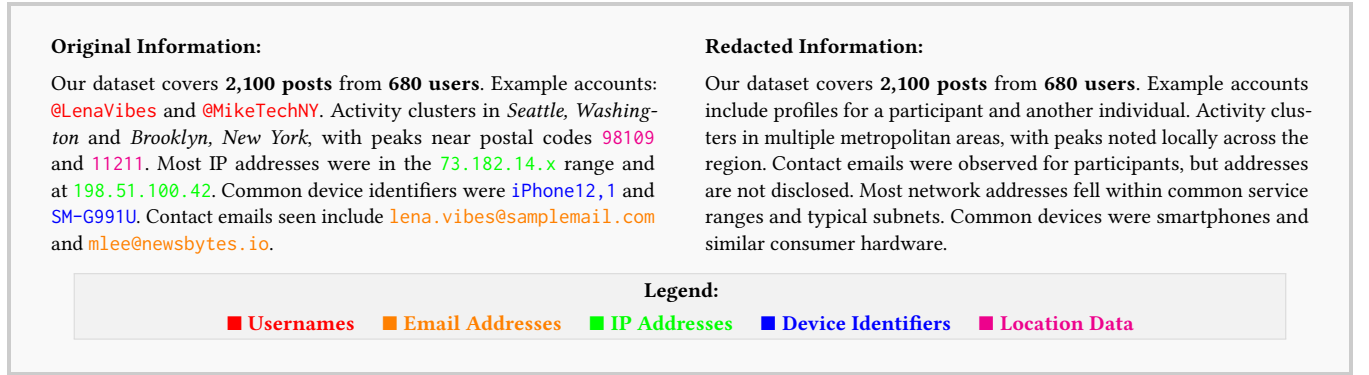

\centering
\begin{mdframed}[
    linewidth=1.5pt,
    linecolor=gray!40,
    backgroundcolor=gray!5,
    roundcorner=8pt,
    innertopmargin=12pt,
    innerbottommargin=12pt,
    innerleftmargin=15pt,
    innerrightmargin=15pt
]
\small

\begin{minipage}[t]{0.48\textwidth}
\textbf{Original Information:}\\[0.5em]
Our dataset covers \textbf{2,100 posts} from \textbf{680 users}. Example accounts: {\color{red}\texttt{@LenaVibes}} and {\color{red}\texttt{@MikeTechNY}}. Activity clusters in \textit{Seattle, Washington} and \textit{Brooklyn, New York}, with peaks near postal codes {\color{magenta}\texttt{98109}} and {\color{magenta}\texttt{11211}}. Most IP addresses were in the {\color{green}\texttt{73.182.14.x}} range and at {\color{green}\texttt{198.51.100.42}}. Common device identifiers were {\color{blue}\texttt{iPhone12,1}} and {\color{blue}\texttt{SM-G991U}}. Contact emails seen include {\color{orange}\texttt{lena.vibes@samplemail.com}} and {\color{orange}\texttt{mlee@newsbytes.io}}.
\end{minipage}%
\hfill
\begin{minipage}[t]{0.48\textwidth}
\textbf{Redacted Information:}\\[0.5em]
Our dataset covers \textbf{2,100 posts} from \textbf{680 users}. Example accounts include profiles for a participant and another individual. Activity clusters in multiple metropolitan areas, with peaks noted locally across the region. Contact emails were observed for participants, but addresses are not disclosed. Most network addresses fell within common service ranges and typical subnets. Common devices were smartphones and similar consumer hardware.
\end{minipage}

\vspace{0.8em}
\centering
\fcolorbox{gray!30}{gray!10}{%
\begin{minipage}{0.95\linewidth}
\centering
\small
\textbf{Legend:} \\[0.3em]
\textcolor{red}{\large$\blacksquare$} \textcolor{red}{\textbf{Usernames}} \quad 
\textcolor{orange}{\large$\blacksquare$} \textcolor{orange}{\textbf{Email Addresses}} \quad 
\textcolor{green}{\large$\blacksquare$} \textcolor{green}{\textbf{IP Addresses}} \quad
\textcolor{blue}{\large$\blacksquare$} \textcolor{blue}{\textbf{Device Identifiers}} \quad 
\textcolor{magenta}{\large$\blacksquare$} \textcolor{magenta}{\textbf{Location Data}}
\end{minipage}%
}
\end{mdframed}
\caption{Comparison of original dataset information (left) with color-coded sensitive data types versus redacted version (right)}
\label{fig:color-coded-info}
\end{figure*}

For each combination of the $6$ domains and $3$ redaction levels, we used an LLM to generate $18$ pieces of original information. An example of information for social media and moderate amount of redaction is shown on the left of Figure \ref{fig:color-coded-info}. It contains $5$ different types of sensitive information, which we have color-coded.

\subsection{Step 2: Information Redaction}
Given the original information, we use an LLM to redact all the sensitive information using a similar approach to prior work, which was validated by human coders and an LLM judge in that work \cite{kaushik2026examiningeffectexplanationsai}. The inputs to this prompt are the original information, scenario description, and sensitive data types, as shown in Figure \ref{fig:system_diagram}. The result of this process is redacted information with all sensitive data removed for each of the $18$ pieces of original information. The redacted information for the example original information in the previous subsection is included on the right of Figure \ref{fig:color-coded-info}.

We verified that each of these $18$ pieces of redacted information did not contain any sensitive data types using manual inspection and an LLM judge that performed with $100\%$ accuracy on this data. 

\subsection{Step 3: Explanation Generation}
After the redaction step, we generate explanations in $5$ different styles: contrastive, general, thorough, normative, and causal, which have been used in other research in explainability (collated and tested in \cite{larasati_effect_2020}). Appendix \ref{sec:appendix-style_descriptions} includes the full instructions given the LLM for each of the explanation styles and explanation examples for the information in Figure \ref{fig:color-coded-info}. A short definition of the styles is included below:
\begin{itemize}
    \item contrastive - difference between the information that was removed and the information that was kept
    \item general - clarity and simplicity not on technical details
    \item thorough - each major step in detail
    \item normative - why the redaction was necessary based on principles or guidelines
    \item causal - specific patterns that caused the redaction, describes the cause-and-effect chain from input to output
\end{itemize}

\begin{figure}[ht]
    \centering
    \includegraphics[width=\columnwidth]{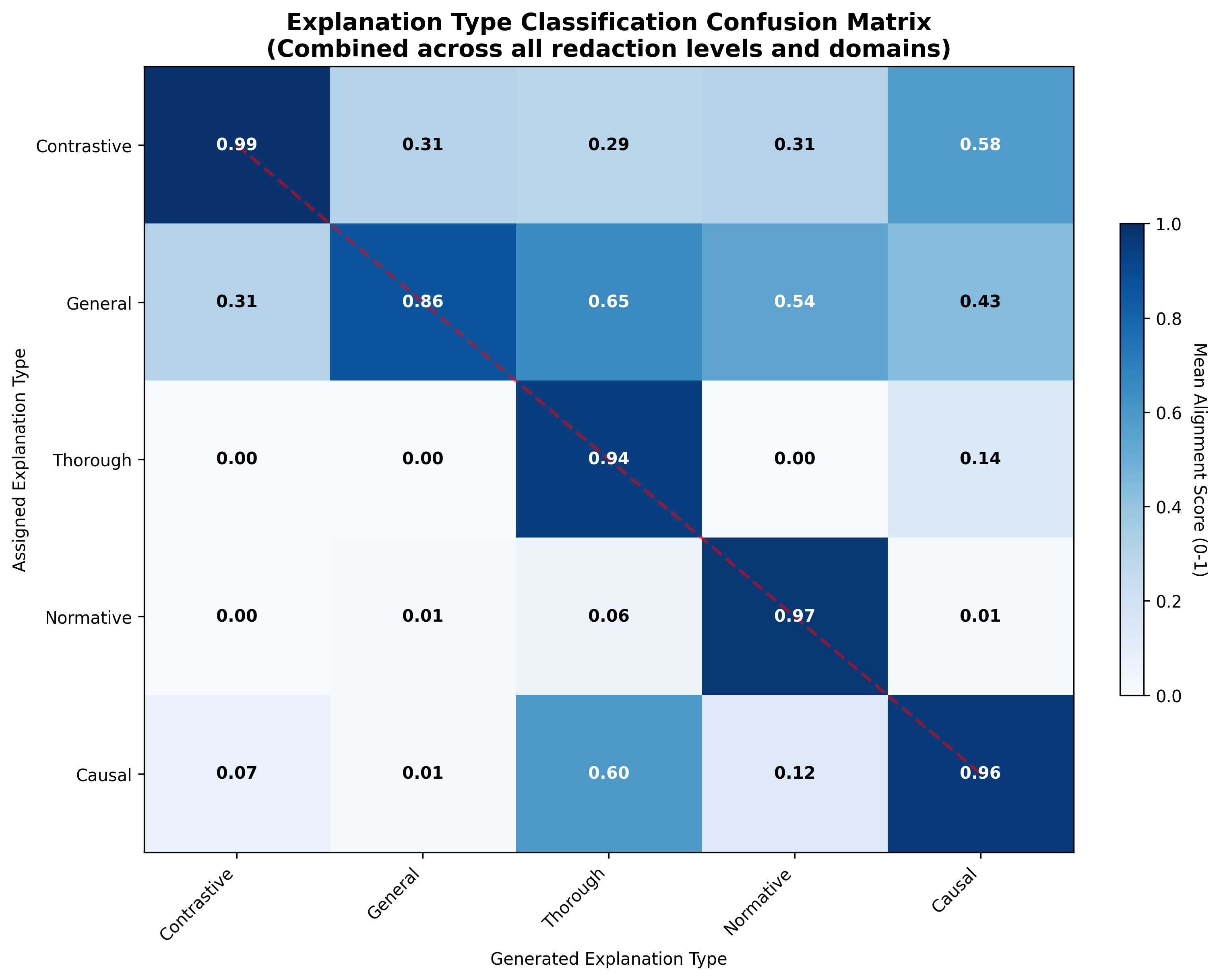}
    \caption{Explanation type confusion matrix between LLM judge evaluation (y-axis) and the generated style (x-axis). The diagonal represents alignment between the generated style and style assigned by the judge.}
    \label{fig:explanation_type_confusion}
\end{figure}

This process results in $5$ explanations, one for each style, for each of the $18$ scenarios. We wanted to verify that these explanations were significantly different from each other (e.g., the contrastive explanation for a scenario was sufficiently different from the thorough explanation for the same scenario). If the generated explanations were too similar, then humans would not be able to distinguish between them and would not have preference differences. For each of the $90$ explanations, we asked an LLM judge to provide a numeric rating $1-5$ to represent how likely it believes that the provided explanation belongs to each of the styles. We then normalized the values to be between $0$ and $1$ and created a confusion matrix to compare the LLM judge's evaluation of the style and the style that was actually generated (Figure \ref{fig:explanation_type_confusion}). As we can see from the figure, the explanation styles are mostly distinct, with the most confusion ($0.65$) occurring between causal and thorough ($1$ on the diagonal indicates alignment between the assigned and generated styles). This result shows us the styles are sufficiently different to proceed to the user study.

\subsection{Step 4: Explanation Redaction}
As seen in the example explanations included in Table \ref{tab:explanation_style_descriptions_examples}, the generated explanations, while being stylistically different, have the potential to leak private information. To address this privacy issue, we used an LLM prompt similar to in Step 2, to remove any sensitive data types present in the explanation. We also inspected these redacted explanations and used an LLM judge to verify that all sensitive information was successfully removed.

\section{User Study Design}
\label{sec:study_design}
To explore the interaction between context, explanation preferences, and trust, we conducted a user study (protocol approved by an internal ethics committee). Participants chose their preferred explanations for different contexts and rated their trust for the scenario. Our two independent variables were domain (medical, education, workplace, social media, financial, and legal), and amount of system redaction (heavy, moderate, light). Our dependent variables were explanation preference and trust. 

\subsection{Participants}
We first performed a power analysis to determine the appropriate number of participants to recruit. We noted that out of the possible $18$ questions we could ask participants for the main part of the study ($6$ domains, $3$ redaction levels), choosing a random $6$ would allow the study to be a reasonable length ($20-30$ min). A sample of around $250$ participants would provide approximately $80\%$ power to detect medium sized interaction effects ($f = 0.20$) in a $6 x 3$ mixed design with $6$ observations per participant at $\alpha = 0.05$. We recruited all our participants from Prolific ($n=249$ as one participant did not complete study appropriately). Detailed demographic information is included in the Appendix in Figure \ref{fig:demographic_pie_charts}.

\subsection{Study Procedure}
\label{sec:study_procedure}
The participants first completed an initial questionnaire containing demographic questions (age, gender, education level) as well as items about their attitudes and experiences with technology and AI. Previous research has found that background can impact trust in automated systems and the effect of explanations on that trust \cite{ehsan_who_2024}, so we asked questions about participants' familiarity, confidence, reliance, and trust in AI systems.

Next participants saw a series of $6$ scenarios, randomly chosen from the $18$ total scenarios. For each scenario, they were first presented with a description of the domain and the sensitive data types that were applicable in that domain. Next, they saw original information as well as the redacted information that was applicable to that domain/redaction level combination.

Participants were then asked to choose which explanation(s) they would have liked the system to include to help explain what the system did for this scenario. They were provided with choices of all $5$ explanation styles, as well as the option to choose ``none of the above.'' After choosing their preferred explanation(s), they completed a series of 5-pt Likert questions intended to evaluate their trust given the scenario and chosen explanation(s). These evaluation items were modified from \cite{hoffman_metrics_2019} and \cite{jian_foundations_2000}, including an explanation satisfaction scale, a trust scale for the explainable AI context, and a checklist for trust between people and automation. These questions included:
\begin{itemize}
    \item The system performed its task accurately and correctly based on the information that is inputted.
    \item The system is reliable and consistent.
    \item I understand what the system is doing and would understand what it would do given different information
    \item The system is agreeable, and I would prefer interacting with this system in place.
    \item I have faith in the ability of the system to perform well in future situations.
    \item I think the system is helpful.
\end{itemize}

After completing the set of questions for each of the $6$ scenarios, participants answered one final question, intended to understand how users would prefer to give feedback about explanations in the future. Our future work includes developing a system that personalizes explanations to the user and context, and we wanted to understand the type of human-in-the-loop feedback users would prefer to provide. Using the framework from \cite{fitzgerald_inquire_2022}, we asked participants about four different interaction types: showing, categorizing, sorting, and evaluating. Specifically, we asked participants to rank their preferences for providing each of these feedback types on a 5-pt Likert scale (only giving them the example sentence not the name of the interaction category):
\begin{itemize}
    \item (Showing) You have to write a sample explanation, so the system understands what kind of explanations you prefer.
    \item (Categorizing) You are provided with an explanation, and you give it a thumbs up or thumbs down depending on how you feel about it.
    \item (Sorting) You are provided with a few explanations, and you have to rank them in order of your preference.
    \item (Evaluating) You are provided an explanation, and you provide a text comment about how you would improve it.
\end{itemize}

\subsection{Research Questions}
We explore a series of research questions to understand the relationships between context, explanation preference, and trust for these privacy-preserving scenarios.

\begin{itemize}
    \item RQ1: How does context impact explanation choice?
    \item RQ2: How is trust affected by context and explanation choice?
    \item RQ3: How do individual users choose explanations?
    \item RQ4: How are user characteristics correlated with explanation choices and trust?
    \item RQ5: How do users want to give feedback on explanations?
\end{itemize}

\section{Results and Discussion}
\label{sec:results}
\subsection{RQ1: How does context impact explanation choice?}
We wanted to understand how domain and redaction level impacted the explanations participants chose. We analyzed this data using ANOVAs paired with Tukey post-hoc tests.

\subsubsection{Domain}
\begin{figure}[ht]
    \centering
    \includegraphics[width=\linewidth]{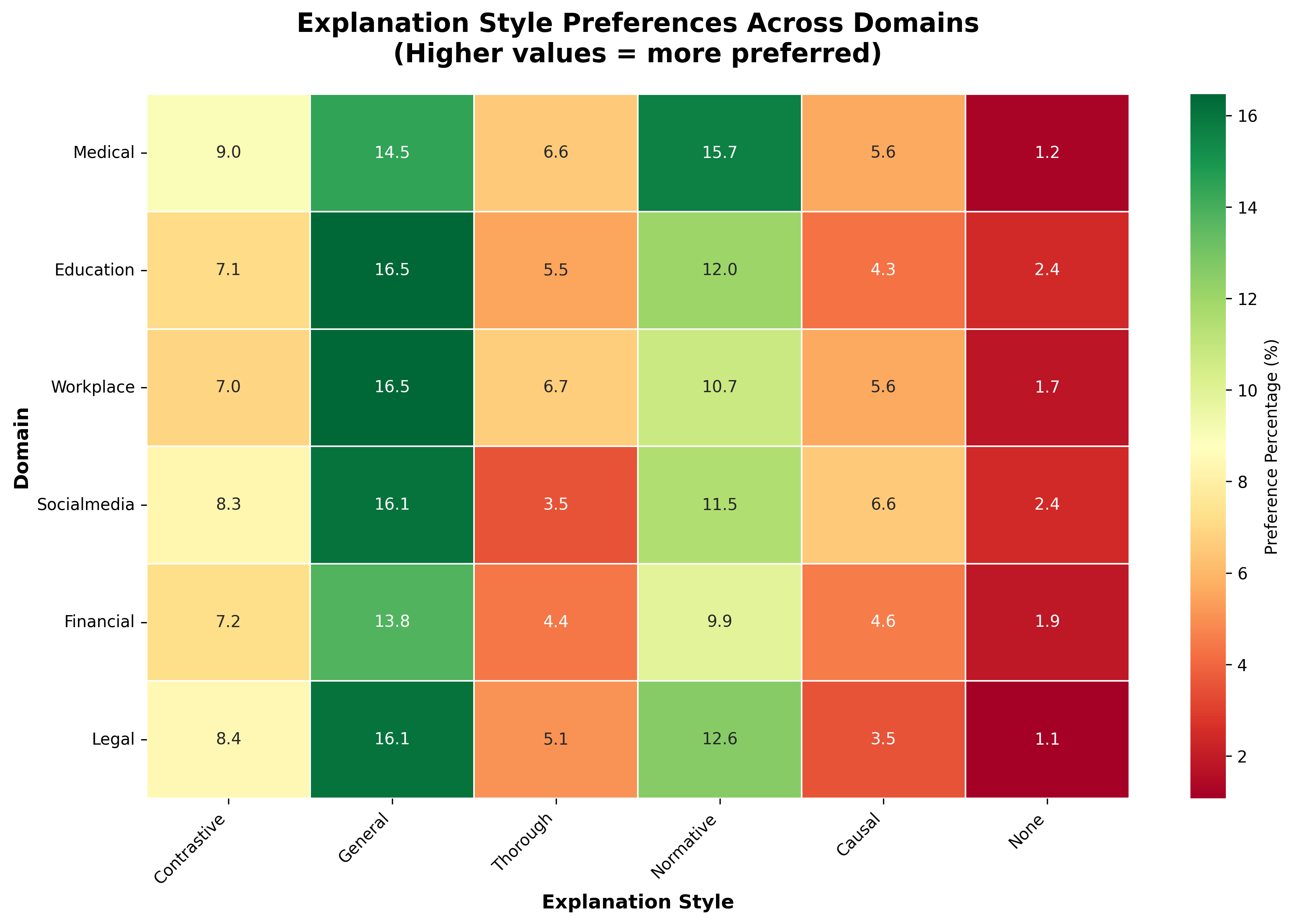}
    \caption{Explanation style preferences across different domains}
    \label{fig:rq1_domain_preferences_heatmap}
\end{figure}

We compared how often participants chose each explanation style for each domain (visualized in Figure \ref{fig:rq1_domain_preferences_heatmap}). We found significant differences for all domains ($p<0.001$), specifically for medical $F(5, 4476) = 29.331$, education $F(5, 4476) = 29.482$, workplace $F(5, 4476) = 26.328$, social media $F(5, 4476) = 27.236$, financial $F(5, 4476) = 22.073$, and legal $F(5, 4476) = 35.174$. Figure \ref{fig:rq1_domain_analysis} in the Appendix shows the results of the post-hoc tests to compute significant differences between the explanation style chosen percentages for each domain, and all domains had many different significant differences. A few interesting findings include:
\begin{itemize}
    \item The medical domain shows a distinct preference distribution, with strong preferences for normative explanations.
    \item General explanations were the most preferred for the other domains, with normative being the second most preferred.
    \item No explanation was preferred the least for all domains.
    \item The preferences for contrastive, thorough, and causal explanations varied between domains, but were preferred less than general and normative.
\end{itemize}

\subsubsection{Redaction Level}
\begin{figure}[ht]
    \centering
    \includegraphics[width=\linewidth]{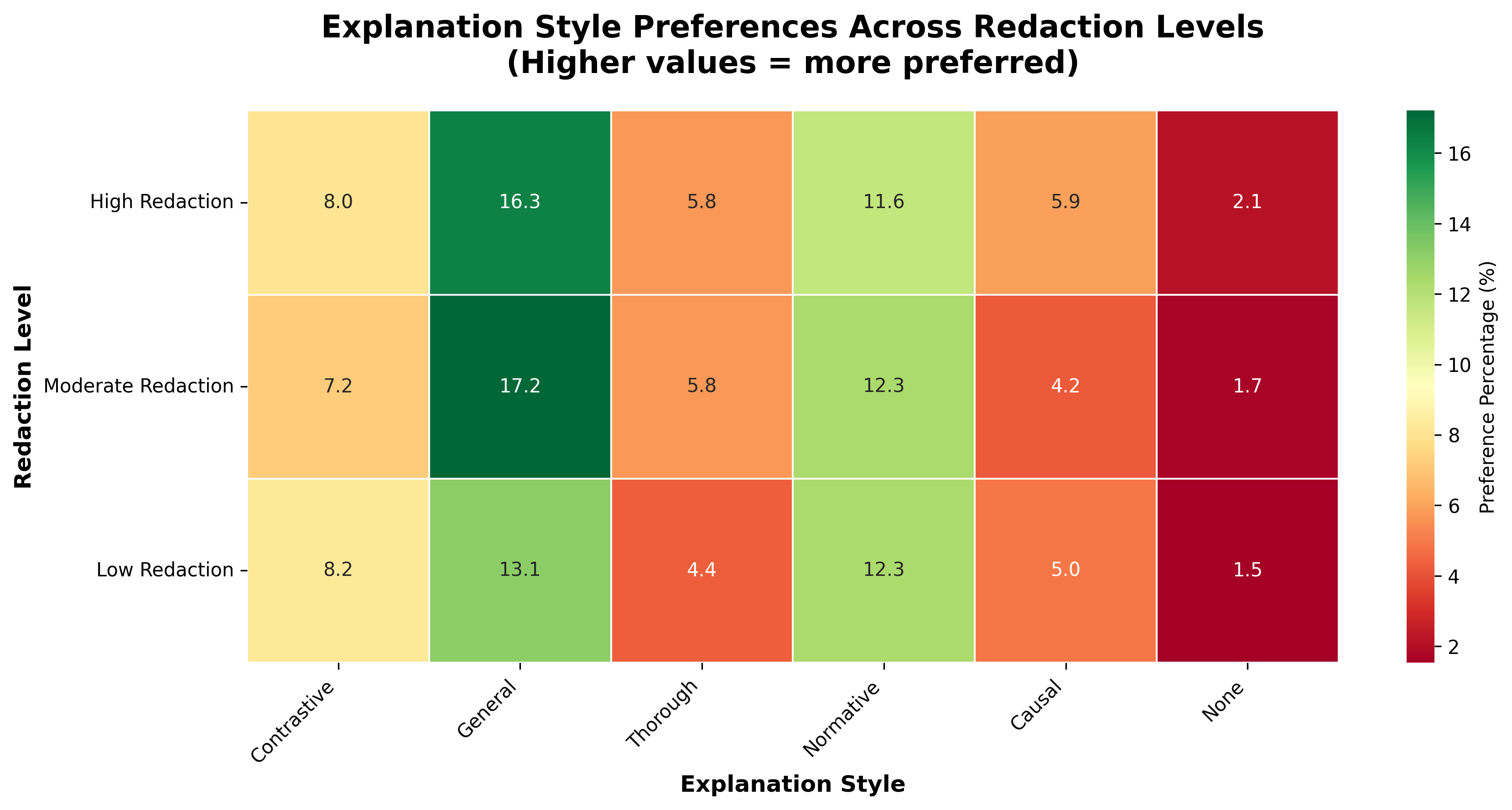}
    \caption{Explanation style preferences across different redaction levels}
    \label{fig:rq1_redaction_preferences_heatmap}
\end{figure}

We compared how often participants chose each explanation style for each redaction level (shown in Figure \ref{fig:rq1_redaction_preferences_heatmap}). We found significant differences for all redaction levels ($p<0.001$), specifically for heavy $F(5, 8958) = 50.71$. moderate $F(5, 8958) = 68.24$, and light $F(5, 8958) = 47.851$ amounts of redaction. Figure \ref{fig:rq1_redaction_analysis} in the Appendix illustrates many different significant pairwise differences which shows that explanation choices vary within a particular amount of redaction. Some interesting findings from these results include:
\begin{itemize}
    \item General explanations were preferred most and no explanation was preferred least for each redaction level.
    \item Thorough was preferred over causal for some redaction levels but not for others.
\end{itemize}

\subsubsection{Interaction between domain and redaction level}
We also analyzed the interaction between these two independent variables to understand how their combination affected participants' explanation preferences. We found that the interaction had a significant effect for general ($F(10, 4464) = 2.324, p < 0.05$) and normative ($F(10, 4464) = 1.911, p < 0.05$) explanations. These styles have preference patterns that depend on both domain and redaction level, and the interaction results are shown in Figure \ref{fig:rq1_interaction_analysis} in the Appendix.

\fbox{\begin{minipage}{0.9\linewidth}
RQ1 Takeaway: Both domain and redaction level as well as their interaction have a significant impact on user explanation preferences.
\end{minipage}}

\subsection{RQ2: How is trust affected by context and explanation choice?}
We wanted to understand how context affected trust as well as how whether users had the explanations of their choice affected trust.

\subsubsection{Context vs. Trust}
We wanted to understand how domain and redaction level impacted the explanations participants chose. We analyzed this data using ANOVAs and Tukey post-hoc tests.

\begin{figure}[ht]
    \centering
    \includegraphics[width=\linewidth]{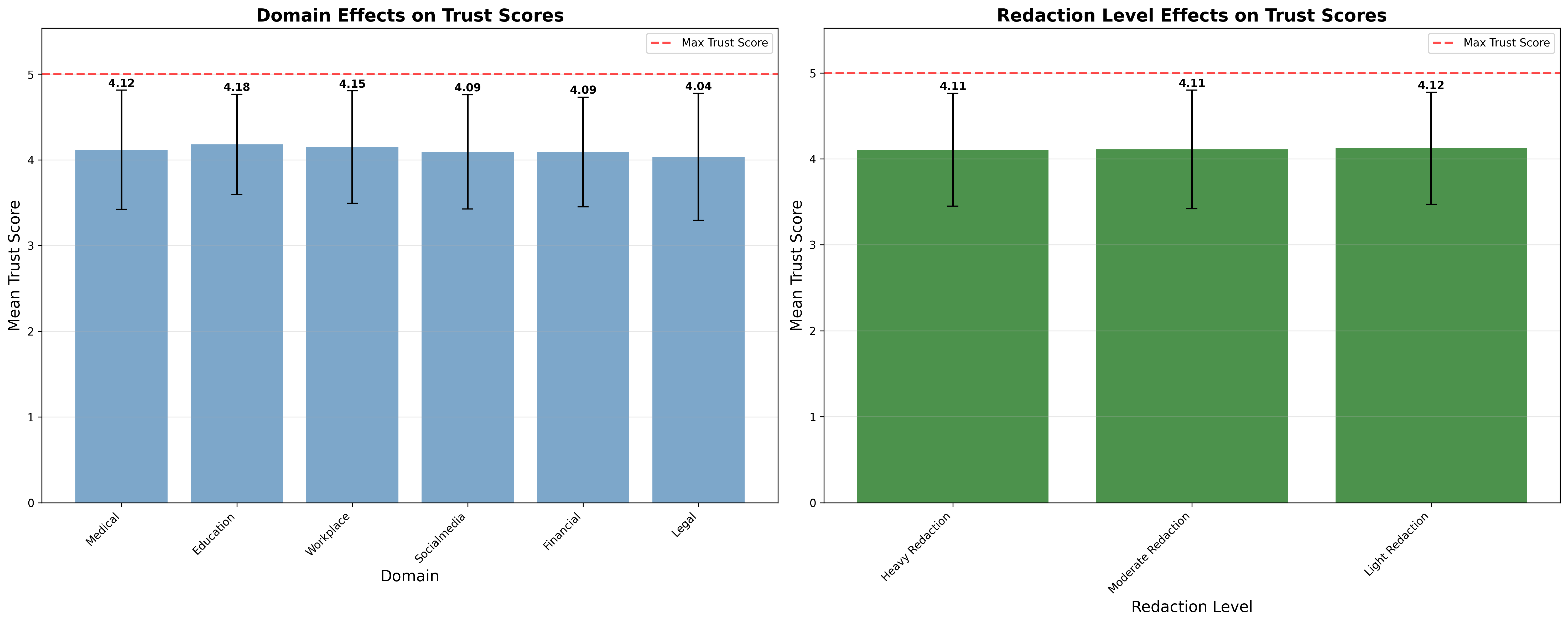}
    \caption{Average trust for different domains (left) and different redaction amounts (right). Maximum trust value is marked in red.}
    \label{fig:rq2_combined_trust_analysis}
\end{figure}

As seen in Figure \ref{fig:rq2_combined_trust_analysis}, trust was relatively high across all domains and redaction levels, and there were no significant differences found. Additionally, there were no significant interaction effects found between these contextual variables when looking at how they affected trust. This tells us that participants generally trusted our system when they were interacting with a scenario with their chosen explanation style.

\fbox{\begin{minipage}{0.9\linewidth}
RQ2 Takeaway: User trust is high and does not vary significantly with context when desired explanation styles are chosen.
\end{minipage}}

\subsubsection{Explanation Choice vs. Trust}
To explore the takeaway further, we compare the trust values from a previous study \cite{kaushik2026examiningeffectexplanationsai} that explored trust using the same Likert-style measures as the current work. In that study, participants were randomly assigned into three conditions: no explanation, general explanations, and thorough explanations. We then grouped the data into three groups:
\begin{itemize}
    \item Group 1: Prior study participants who did not receive explanations
    \item Group 2: Prior study participants who received an explanation randomly (general or thorough)
    \item Group 3: Current study participants who were able to chose their desired explanation(s)
\end{itemize}

\begin{figure}[ht]
    \centering
    \includegraphics[width=\linewidth]{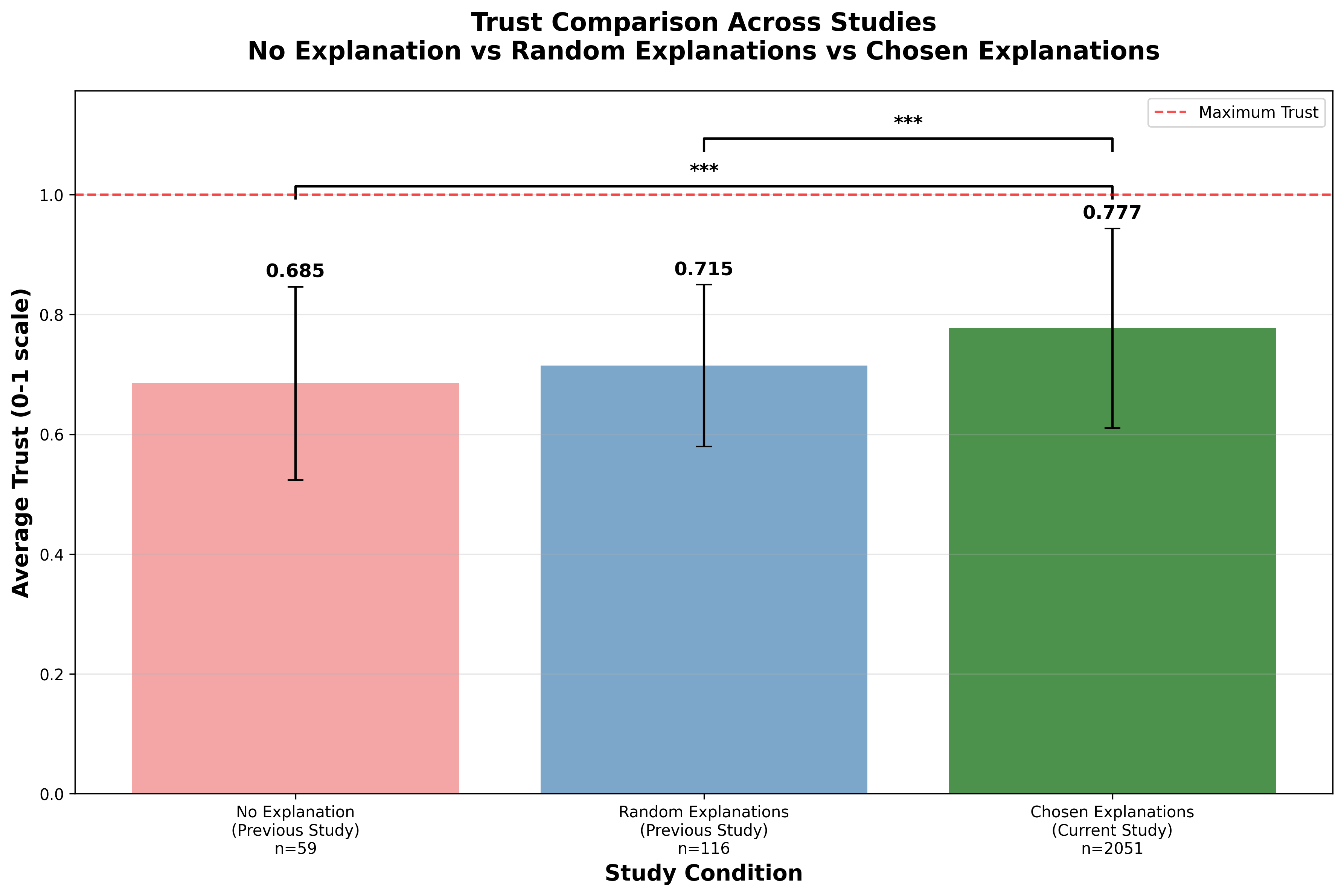}
    \caption{Trust comparison across studies with significant differences marked with three asterices ($p<0.001$). No explanation and random explanation from a previous study is compared to chosen explanation from this study}
    \label{fig:rq2_three_group_comparison}
\end{figure}

We compared average trust across these three groups in Figure \ref{fig:rq2_three_group_comparison} and conducted an ANOVA and Tukey post-hoc tests. We found significant differences between groups ($F(2, 2223) = 16.1901, p<0.001$) as well as pairwise differences between Group 3 and Groups 1/2. When participants were given their chosen explanation, they trusted our system more than no explanation or a random explanation.

\fbox{\begin{minipage}{0.9\linewidth}
RQ2 Takeaway: When users are given their chosen explanation, their trust is higher than a random or no explanation.
\end{minipage}}

\subsection{RQ3: How do individual users choose explanations?}
To understand how participants chose explanations, we analyzed both the context dependence of participants and how often participants preferred two or more explanation styles equally for the same context.

\subsubsection{Context Dependence}
We believed that participant explanation preferences would be dependent on context, and we have support for that hypothesis using overall trends as seen in the analysis of RQ1. However, we also wanted to analyze individual participant explanation preferences and understand how their preferences might be dependent on the two different contextual variables in this study: domain and redaction level.

To quantify how much a participant's explanation choices vary across contexts, we compute a \textbf{context dependence score} that captures the extent to which their preferences change across domains (or redaction levels). Intuitively, this score measures the \textbf{relative variability in how often each explanation style is selected across contexts}, normalized so that $0$ indicates consistent preferences and $1$ indicates maximal variation. In other words, it reflects how much a participant's explanation choices shift across contexts, with higher values indicating stronger context dependence.

To compute this score, we follow these steps (explained here for domain context dependence; the same process is used for redaction level):
\begin{enumerate}
    \item Construct a list of (domain, redaction level, chosen style) tuples for each participant. If multiple explanations are selected in a single condition, multiple tuples are recorded.
    \item Group choices by domain and count how many times each explanation style is selected within each domain.
    \item Convert counts to proportions by dividing by the total number of choices in that domain. For example, if the medical domain appears $5$ times and a participant selects a thorough explanation $2$ times, the proportion is $2/5 = 0.4$.
    \item For each explanation style, collect its proportions across all domains.
    \item Compute the \textbf{coefficient of variation} (CV), defined as the standard deviation divided by the mean, to capture relative variability across domains.
    \item Normalize the CV by its theoretical maximum, $\sqrt{N - 1}$, where $N$ is the number of domains.
    \item Average the normalized CV values across explanation styles to obtain a single context dependence score in the range $[0, 1]$, where $0$ indicates consistent preferences and $1$ indicates maximal variability.
\end{enumerate}

\begin{figure}
    \centering
    \includegraphics[width=\linewidth]{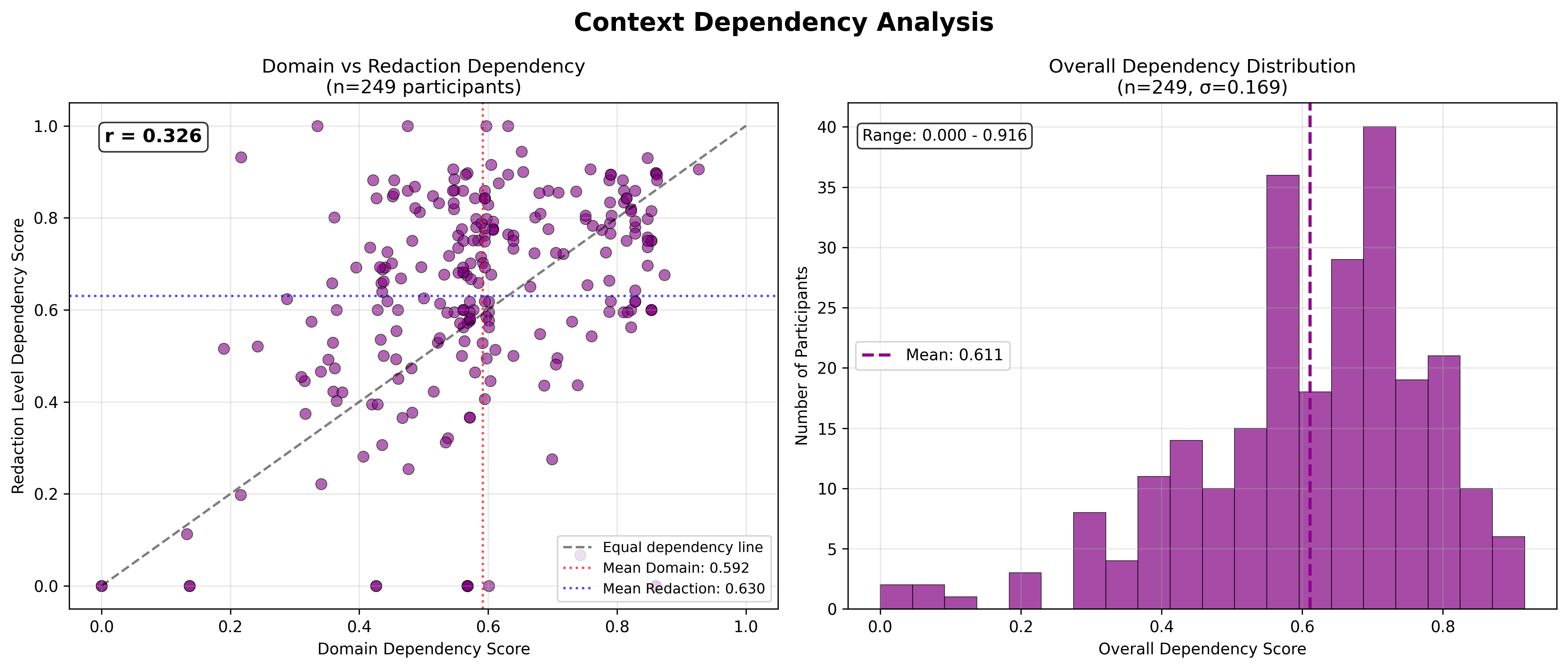}
    \caption{(Left) Domain and redaction level context dependency scores for each participant. (Right) Distribution of overall context dependence of participants (average of domain and redaction level scores)}
    \label{fig:rq3_context_dependency_histograms}
\end{figure}

We performed this process for both the domain and redaction level variable for every participant and then averaged these two values together to obtain an overall context dependency score for that participant. Figure \ref{fig:rq3_context_dependency_histograms} visualizes the domain and redaction level contextual dependency scores for each participant on the left plot. As we can see from the wide distribution of points, participants had a wide variation in their context dependency, with the two variables only weakly correlated ($r = 0.236$).

We also plot the distribution of overall dependency scores, and the average dependency was $0.611$. The high standard deviation of $0.169$ indicates a large variability between participants. This variability indicates that some participants may not change their explanation preferences based on context, and others may have preferences greatly dependent on context.

\fbox{\begin{minipage}{0.9\linewidth}
RQ3 Takeaway: Users have varying levels of dependence on context when choosing their preferred explanation style.
\end{minipage}}

\subsubsection{Multiple Explanation Styles for a Particular Context}
In Figure \ref{fig:explanation_type_confusion}, we saw that an LLM judge generally evaluated the explanation styles as distinct from each other. Since we gave participants the option to choose multiple, equally preferred explanation styles for a given context, we wanted to understand how many times participants utilized this option. Choosing multiple styles could indicate that the participant saw the explanations as similar, but it could also indicate that the participant viewed these styles as equally preferred.

\begin{figure}
    \centering
    \includegraphics[width=\linewidth]{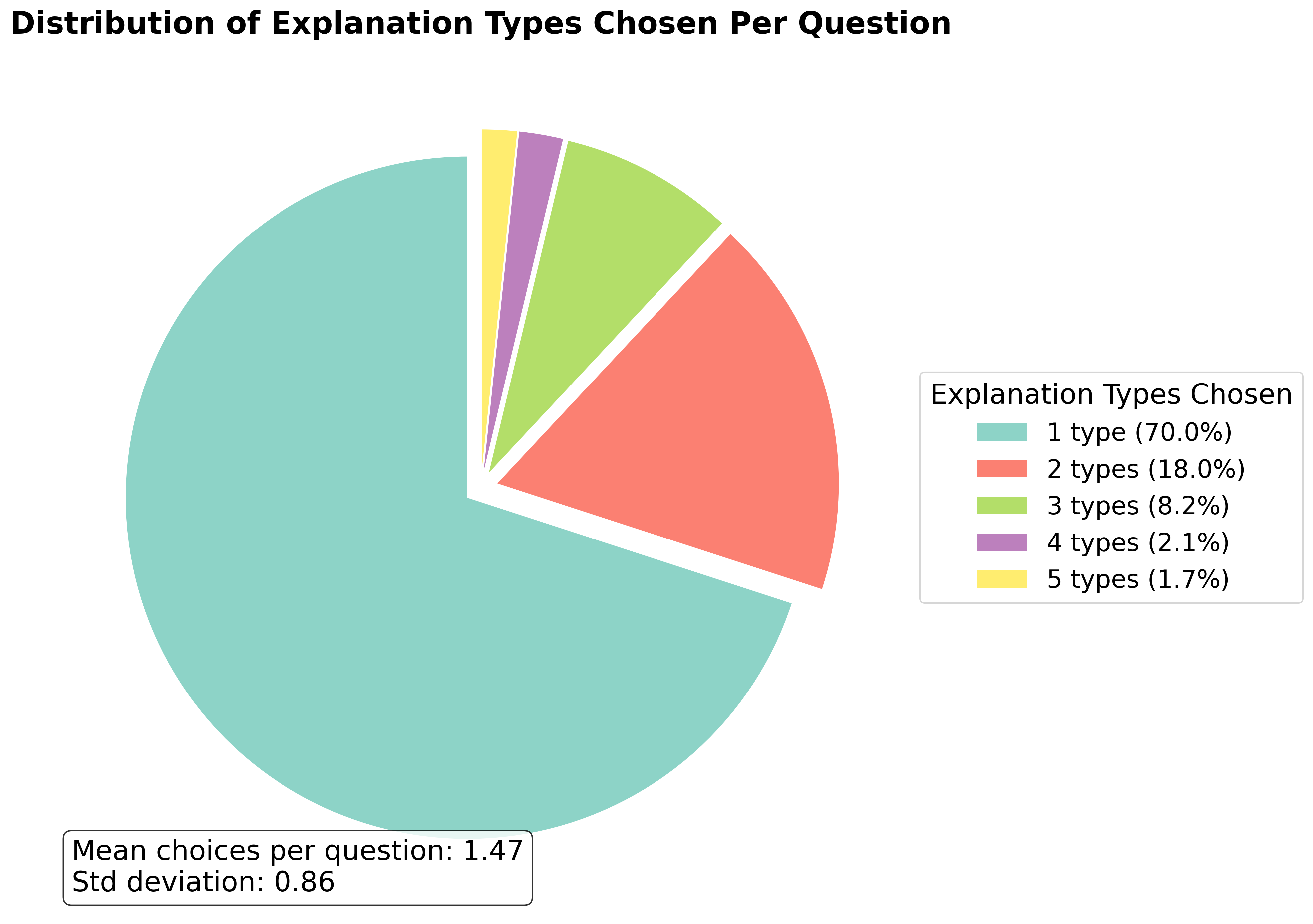}
    \caption{The number of times participants chose multiple explanation styles for a single context. The times that participants chose $1$ style per context, $2$ styles per context, etc. are shown in the chart.}
    \label{fig:rq3_explanation_choices_per_question}
\end{figure}

Figure \ref{fig:rq3_explanation_choices_per_question} shows this distribution with participants choosing one style per context $70\%$ of the time and two styles per context $18\%$ of the time. It was relatively uncommon for participants to choose more than $2$ styles per given context. This indicates that participants had strong opinions most of the time for which style they would prefer for a given scenario, and it supports our hypothesis that choosing multiple styles is more likely due to participants having similar preferences for a few styles in a few contexts rather than seeing styles as overall similar (which would have resulted in much higher occurrence of multiple choices per context).

\begin{figure}
    \centering
    \includegraphics[width=\linewidth]{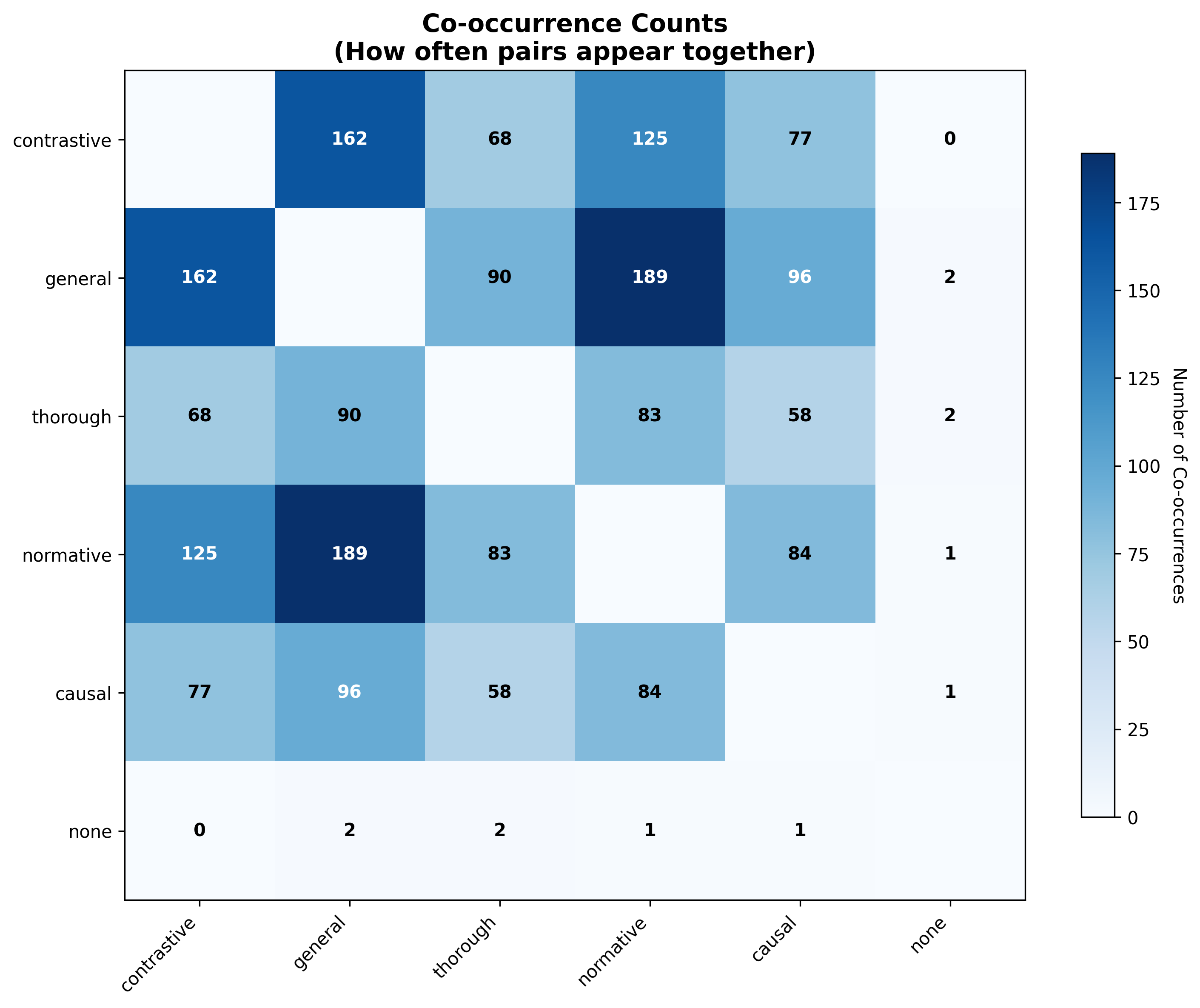}
    \caption{Matrix indicating how often two explanation style were chosen together for the same scenario}
    \label{fig:rq3_explanation_cooccurrence}
\end{figure}

We also performed a co-occurrence analysis to understand how users combined different explanation styles in their responses. It can show us whether certain explanation types are frequently chosen together or tend to be mutually exclusive. The results are shown in Figure \ref{fig:rq3_explanation_cooccurrence}. For example, general explanations and normative explanations were chosen together more frequently than other combinations, which could indicate the participant preferences for them were similar. In contrast, thorough and causal explanations were chosen together less frequently, indicating that preferences for those were dissimilar. 

\fbox{\begin{minipage}{0.9\linewidth}
RQ3 Takeaway: Users have relatively strong opinions about explanation styles for a particular context, and there is not a strong co-occurence between different styles.
\end{minipage}}

\subsection{RQ4: How are user characteristics correlated with explanation choices and trust?}
As explained in Section \ref{sec:study_procedure}, we asked participants their age, gender, education level, and their attitudes towards AI and technology. We want to understand how these user characteristics are correlated with the explanation preferences participants chose and their average trust in our system.

\subsubsection{Explanation Preferences}
\begin{figure}
    \centering
    \includegraphics[width=\linewidth]{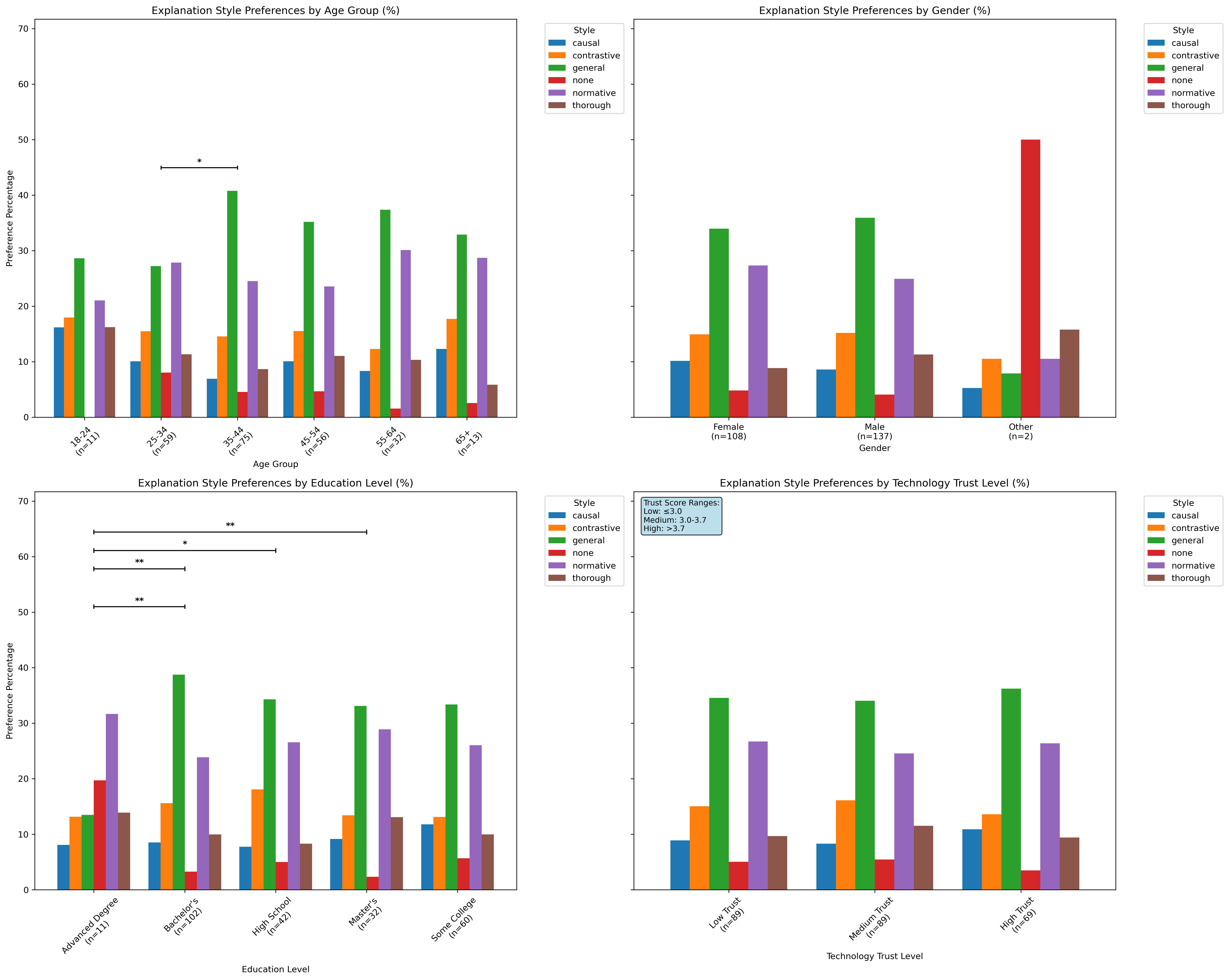}
    \caption{Explanation Preferences for different user characteristics groups: age, gender, education level, trust in technology. Significant pairwise differences are marked with one, two, or three asterisks ($p<0.05$, $p<0.01$, and $p<0.001$, respectively)}
    \label{fig:rq4_new_preferences_analysis}
\end{figure}

We first looked at the explanation choices of participants with different user characteristics, pictured in Figure \ref{fig:rq4_new_preferences_analysis}. We performed ANOVA tests paired with Tukey post-hoc tests to see if there were any significant differences between groups. We found significance with the frequency of choosing general explanations between different age groups ($F(5, 240) = 2.4530, p < 0.05$). Specifically, those aged $35-44$ preferred general explanations more those $25-34$.

We also found significance with the frequency of choosing general explanations between different education levels ($F(4, 242) = 3.1128, p < 0.05$), where those with an advanced degree chose general explanations less than those with a bachelor's degree. Additionally, there was significance in the frequency of choosing no explanation between different education levels ($F(4, 242) = 3.3221, p < 0.05$). Participants with an advanced degree chose no explanation much less than those with high school, some college, bachelor's, or masters education. We did not find significant differences for gender or trust in technology.

\subsubsection{Trust in our System}
\begin{figure}
    \centering
    \includegraphics[width=\linewidth]{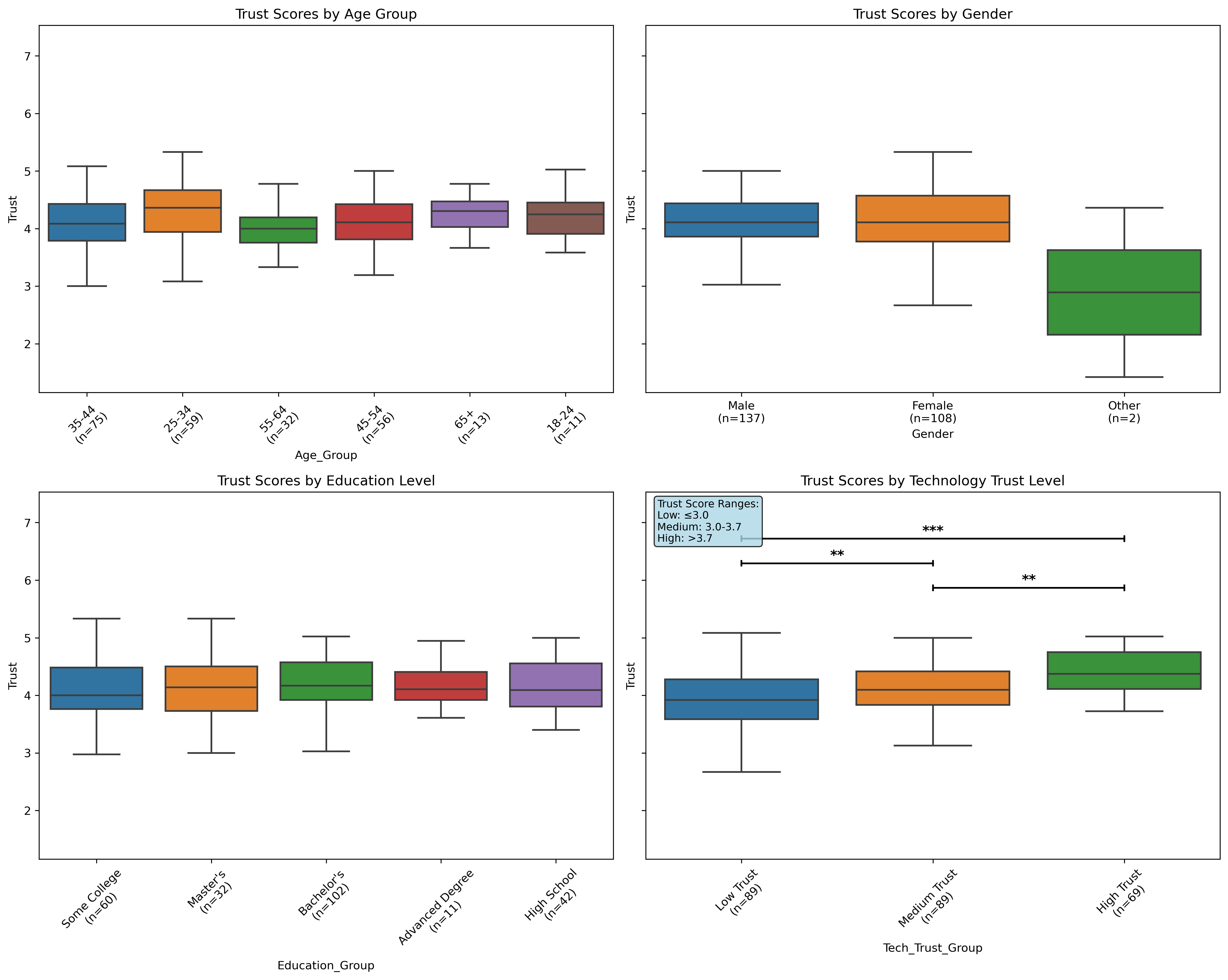}
    \caption{Trust in our system for different user characteristics groups: age, gender, education level, trust in technology. Significant pairwise differences are marked with one, two, or three asterisks ($p<0.05$, $p<0.01$, and $p<0.001$, respectively)}
    \label{fig:rq4_new_trust_analysis}
\end{figure}

We then examined how average trust in our system was correlated with user characteristics, shown in Figure \ref{fig:rq4_new_trust_analysis}. We found one major significant result in this analysis, the participants' prior trust in AI impacted their trust in our system significantly ($F(2, 244) = 18.3338, p < 0.05$). Participants who trusted AI and technology more also trusted our system more, which aligns with results found in prior work \cite{kaushik2026examiningeffectexplanationsai}.

\fbox{\begin{minipage}{0.9\linewidth}
RQ4 Takeaway: Age and education level have a limited correlation with explanation style preference, and a higher baseline trust in AI and technology is associated with a higher trust in our system.
\end{minipage}}

\subsection{RQ5: How do users want to give feedback on explanations?}
As we have seen from the results of the previous research questions, the preferred explanation style for a particular user for a particular context is not straightforward to infer. Individual and contextual differences interact in complex ways, so any system whose role is to choose the correct explanation for a user for a context would need some kind of user feedback to determine whether chosen explanations are approved by the user. We use the framework presented in \cite{fitzgerald_inquire_2022} and asked the question included in Section \ref{sec:study_procedure} to understand which feedback participants preferred.

\begin{figure}
    \centering
    \includegraphics[width=\linewidth]{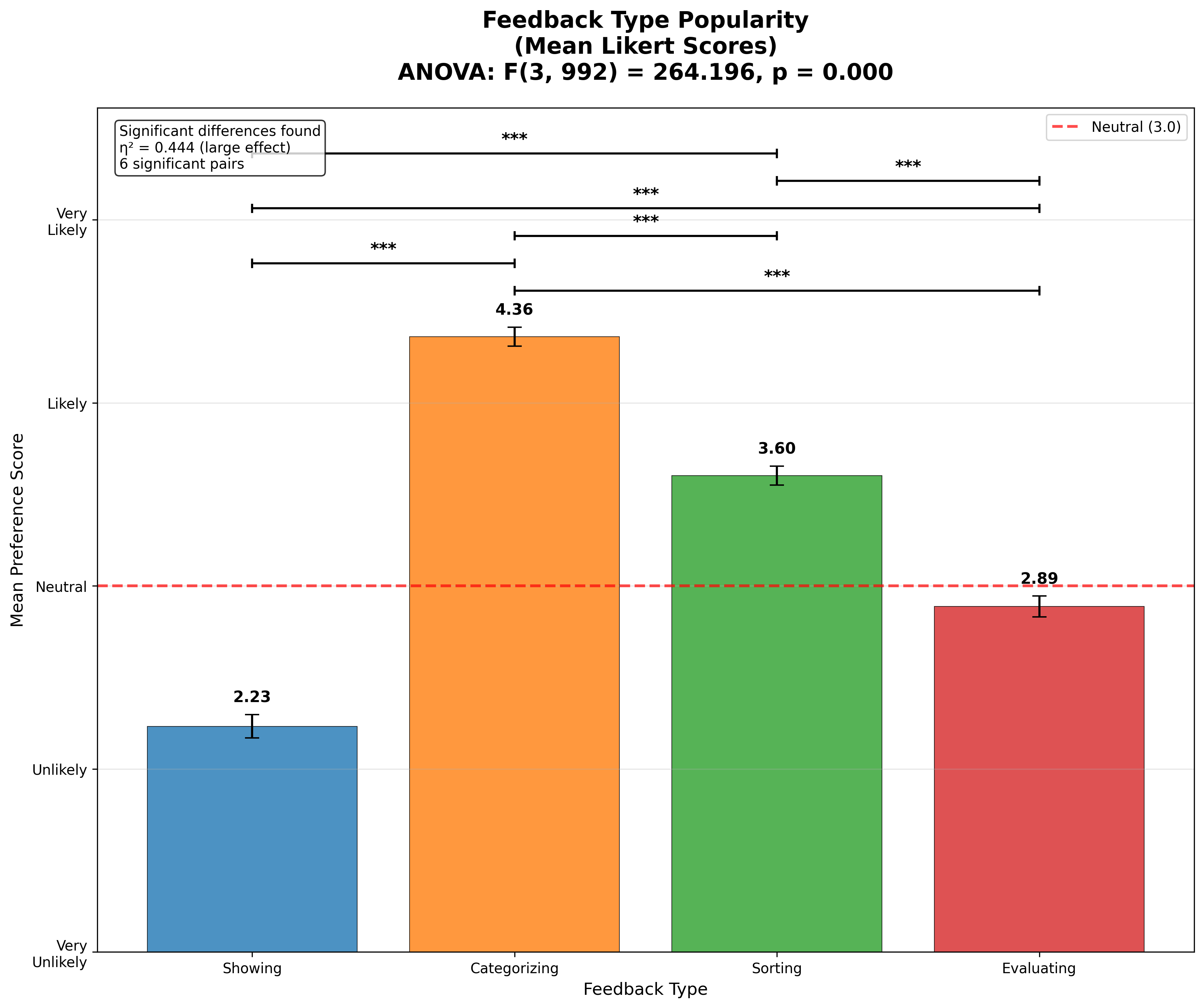}
    \caption{Preferred feedback types by participants for to help a system learn their explanation preferences}
    \label{fig:rq5_feedback_popularity}
\end{figure}

Figure \ref{fig:rq5_feedback_popularity} visualizes user preferences for the different feedback types, where we found significant differences between types ($F(3, 992) = 264.196, p<0.001$). Specifically, the order participants preferred the types from most preferred to least was: categorizing (thumbs up or down), sorting (ranking a few choices), evaluating (writing how an explanation could be improved), and showing (writing a sample explanation). These results are consistent with participants preferring feedback types with the least cognitive load. Any future work in developing a system to provide personalized explanations to users could use this result to inform the kinds of user feedback the system could elicit to improve its explanation choices.

\fbox{\begin{minipage}{0.9\linewidth}
RQ5 Takeaway: Users prefer categorizing and sorting feedback types when providing feedback about explanations generated by a system.
\end{minipage}}

\section{Conclusion}
\label{sec:conclusion}
AI mediators are increasingly used to filter, transform, and redact information in human interactions, which requires users to interpret the transformed outputs. In this work, we examined how users select and engage with explanations in such scenarios, focusing on the roles of context, trust, and individual differences. Our findings show that explanation preferences are not fixed, but vary systematically with domain and level of information redaction, highlighting the limitations of choosing a single explanation type for all situations. We further find that users' trust in AI-mediated systems is closely tied to whether they receive their preferred explanations, which suggests that a system should adapt to user preference for appropriate trust calibration. Additionally, each user exhibited different explanation preferences and dependence on context, indicating that preferences are shaped by the combination of situational and user-specific factors. Finally, users expressed clear expectations for how they wanted to provide feedback on explanations provided by the mediator, highlighting the importance of mechanisms that allow systems to adapt to user feedback.

These findings point to several implications for the design of AI-mediated interactions. Instead of presenting a fixed or uniform explanation, systems should \textbf{adapt explanation strategies to both context and user characteristics}, selecting the most appropriate explanation for a given situation. Our results show that explanation preferences very systematically with variables such as domain and the level of information redaction, which means explanation design should be \textbf{context-aware}. The presence of individual differences indicates that systems could benefit from personalization, learning which types of explanations are most effective for different users over time. Finally, user feedback mechanisms should be treated as a core component of the interaction and can enable systems to iteratively refine explanation selection in privacy-sensitive scenarios. Together, these findings highlight the need for AI-mediated interfaces that are adaptive, personalized, and responsive to both contextual changes and user needs.

Although this work provides insight into how users engage with explanations in AI-mediated privacy-sensitive scenarios, it has several limitations that lead to future research. First, our study examines a controlled, discrete set of domains and redaction scenarios, which do not fully capture the complexity of real-world applications where information constraints and user goals are more dynamic. Second, our findings are based on short-term interactions, and future work is needed to understand how explanation preferences and trust evolve long-term. Additionally, while we identify patterns in how context and individual differences shape explanation preferences, translating these insights into actual systems adapting to users requires further exploration into how such systems can reliably learn and update user models over time.

As AI mediators become increasingly used in domains such as healthcare and finance, the challenge of balancing privacy, transparency, and trust will become even more critical. Our work suggests that addressing this challenge requires moving beyond static explanations and towards systems that can dynamically adapt to individual users and changing contexts. We hope that this work motivates future research on designing AI-mediated interactions that are privacy-preserving, trustworthy, and responsive to the needs of users.





\bibliographystyle{ACM-Reference-Format}
\bibliography{bibliography}

\appendix
\section{Appendix}
\subsection{Domain Information}
\label{sec:appendix-domain_information}
Table \ref{tab:domain_information} includes the name, two parties interacting, domain description and sensitive data types for each of the $6$ domains explored in this work.

\begin{table*}[ht]
\centering
\caption{Information defining the $6$ domains explored in this work}
\label{tab:domain_information}
\begin{tblr}{Q[1cm]Q[2.75cm]Q[2.75cm]XX} \hline
\textbf{Domain Name} & \textbf{Party A} & \textbf{Party B} & \textbf{Description} & \textbf{Sensitive Data Types} \\ \hline
medical \cite{aitken_public_2016, jones_public_2022} & clinical researcher & external data analyst & A clinical researcher is collaborating with an external data analyst to analyze patient data for a medical study. They need to share aggregated patient information while ensuring compliance with privacy regulations. & patient names, dates of birth, home addresses, patient IDs, medical record numbers, vitals (e.g., specific blood pressure or heart rate values), diagnosis information (e.g., specific conditions like diabetes or hypertension)\\
education \cite{slade_learning_2013, regan_ethical_2019} & university researcher & policy advisor & A university researcher is working with a policy advisor to analyze educational data. They need to share aggregated student information while ensuring compliance with privacy regulations. & student names, social security numbers, grades, student IDs, home addresses, phone numbers, email addresses\\
workplace \cite{kellogg_algorithms_2020, teebken_privacy_2021} & HR manager & external consultant & An HR manager is collaborating with an external consultant to analyze employee data. They need to share aggregated employee information while ensuring compliance with privacy regulations. & employee names, salaries, employee IDs, performance reviews, home addresses, phone numbers, emergency contacts\\
social media \cite{singhal_toward_2024, saravanakumar_privacy_2016} & researcher & journalist & A researcher is collaborating with a journalist to analyze social media data. They need to share aggregated user information while ensuring compliance with privacy regulations. & usernames, real names, email addresses, phone numbers, location data, IP addresses, device identifiers\\
financial \cite{abbe_privacy-preserving_2012, el_haddad_understanding_2018} & financial analyst & regulatory auditor & A financial analyst is working with a regulatory auditor to analyze financial data. They need to share aggregated financial information while ensuring compliance with privacy regulations. & account numbers, social security numbers, transaction details, credit scores, income information, bank routing numbers, loan application details\\
legal \cite{silkenat_privacy_2014} & legal researcher & academic collaborator & A legal researcher is collaborating with an academic collaborator to analyze legal data. They need to share aggregated legal information while ensuring compliance with privacy regulations. & client names, case numbers, litigation tactics (e.g., motions to dismiss, discovery strategies), settlement amounts, court dates, attorney-client communications, witness information \\ \hline
\end{tblr}
\end{table*}

\subsection{Explanation Style Descriptions and Examples}
\label{sec:appendix-style_descriptions}
Table \ref{tab:explanation_style_descriptions_examples} includes the instructions fed into the LLM prompt to generate explanations in each of the $5$ explanation styles as well as an example of a generated explanation for each style for the social media / moderate redaction level case.

\begin{table*}[ht]
\centering
\caption{The prompt instructions and and example for each of the $5$ explanation styles}
\begin{tblr}{Q[1.5cm]XX}
\hline
\textbf{Style Name} & \textbf{Prompt Instructions}  & \textbf{Example}  \\ \hline
contrastive         & Provide a contrastive explanation of the redaction. Focus on the difference between the information that was removed and the information that was kept. Explain why the system decided to redact this particular content rather than leaving it unchanged or redacting something else. Use comparative phrasing like ``instead of,'' ``rather than,'' or ``because unlike.'' Emphasize reasoning through contrast — not general summaries or technical steps.                                                                                                                          & We kept aggregate counts because unlike usernames, emails, IPs, and device IDs, they don’t identify individuals. We referenced accounts and devices generically instead of listing specific handles or models. Locations were broadened to metro areas rather than naming cities or postal codes. Network info was described as common ranges instead of specific addresses to reduce re-identification risk.                                                                         \\
general             & Provide a general explanation of the redaction. Use plain, straightforward language to describe what kinds of information were removed and why. Focus on clarity, simplicity, and broad understanding — not on technical system processes or legal justifications. Avoid complex phrasing or detail. The explanation should sound like something you’d tell a non-technical reader who just wants to understand the main point.                                                                                                                                                  & We removed details that could identify people or devices. Specific usernames, emails, exact cities and ZIP codes, IP addresses, and device models were replaced with generic terms and broader regions. We kept overall counts and patterns. This limits the chance of tracing data back to an individual while still showing trends.                                                                                                                                                 \\
thorough            & Provide a thorough explanation of the redaction. Describe each major step the system took - from detecting sensitive elements to removing or replacing them - in the order they occurred. Explain how the system ensured that no identifiable data remained. Include details about detection, categorization, and replacement logic. Avoid simplifications or value-based justification; focus on complete procedural detail.                                                                                                                                                    & 1) Parsed text; matched handles (@...), emails, IPs, postal codes, device models, and city/state names. 2) Tagged as usernames, emails, IPs, locations, device IDs. 3) Generalized: accounts to participants; cities to metropolitan areas; zip/IP to non-specific ranges; emails to undisclosed; devices to smartphones. 4) Validated with a second pass: no identifiable tokens remaining; counts preserved; confirmed no phone numbers. \\
normative           & Provide a normative explanation of the redaction. Focus on why the redaction was the correct or necessary action based on privacy principles, ethical guidelines, or legal compliance (for example, protecting confidentiality or following data protection laws). Explain the values or obligations that justify removing the information, not the technical method. Avoid step-by-step or causal descriptions — this explanation should sound principled and justificatory.                                                                                                    & Specific usernames, emails, IPs, precise ZIPs, and device identifiers were converted into generic, aggregated descriptions to prevent singling out and re‑identification. This applies data minimization and purpose limitation, protects confidentiality of social media users, and supports lawful, proportionate sharing under data protection obligations while preserving aggregate analytical value.                                                                            \\
causal              & Focus on identifying the specific signals, patterns, or detection rules that caused the system to redact the text. Describe the cause-and-effect chain leading from input (e.g., presence of a person's name) to output (e.g., removal or generalization). Emphasize mechanistic causation rather than normative or comparative reasoning. The goal is to clarify the triggers and logic that directly led to the redaction decision. Tone: Mechanistic, explanatory, technical but concise. Model should focus on: cause-effect logic and specific triggers of system behavior. & Detections caused redaction: @handle pattern flagged usernames (@LenaVibes, @MikeTechNY) to generalized to ``participant''. Email format local@domain (lena.vibes@samplemail.com, mlee@newsbytes.io) to removed. City names and 5-digit ZIPs (98109, 11211) to broadened to regions. IPv4 strings (73.182.14.x, 198.51.100.42) to replaced by service ranges. Device model codes (iPhone12,1, SM-G991U) to labeled as smartphones.                           \\ \hline
\end{tblr}
\label{tab:explanation_style_descriptions_examples}
\end{table*}

\subsection{Detailed Participant Demographics}
\begin{figure*}[ht]
    \centering
    \includegraphics[width=\linewidth]{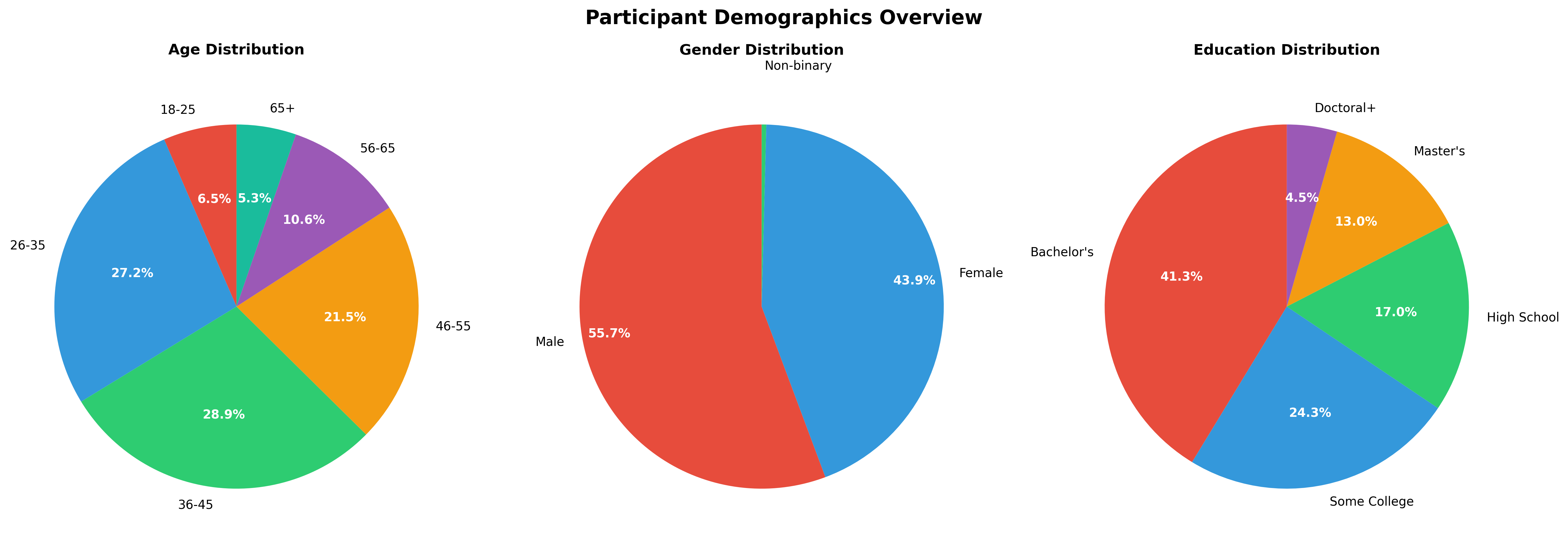}
    \caption{Detailed age, gender, and education distribution of participants}
    \label{fig:demographic_pie_charts}
\end{figure*}

Figure \ref{fig:demographic_pie_charts} shows the age, gender, and education distribution of study participants.

\subsection{Domain and Redaction Level Impact on Explanation Choice}

\begin{figure*}[ht]
    \centering
    \includegraphics[width=0.75\linewidth]{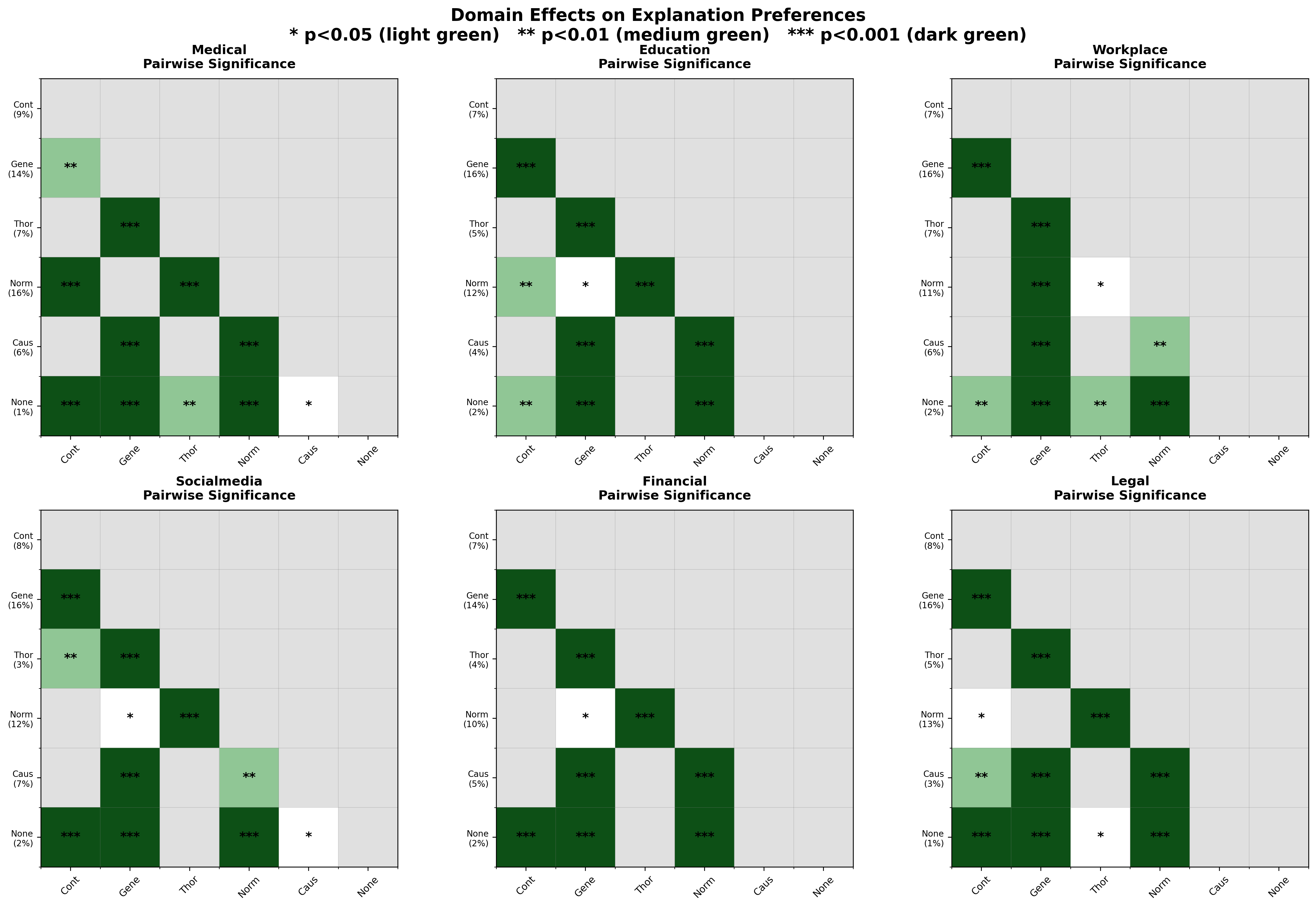}
    \caption{Pairwise comparison of explanation styles chosen for each domain. Significant differences are marked in light green, medium green, and dark green for $p<0.05$, $p<0.01$, and $p<0.001$, respectively.}
    \label{fig:rq1_domain_analysis}
\end{figure*}

\begin{figure*}[ht]
    \centering
    \includegraphics[width=0.75\linewidth]{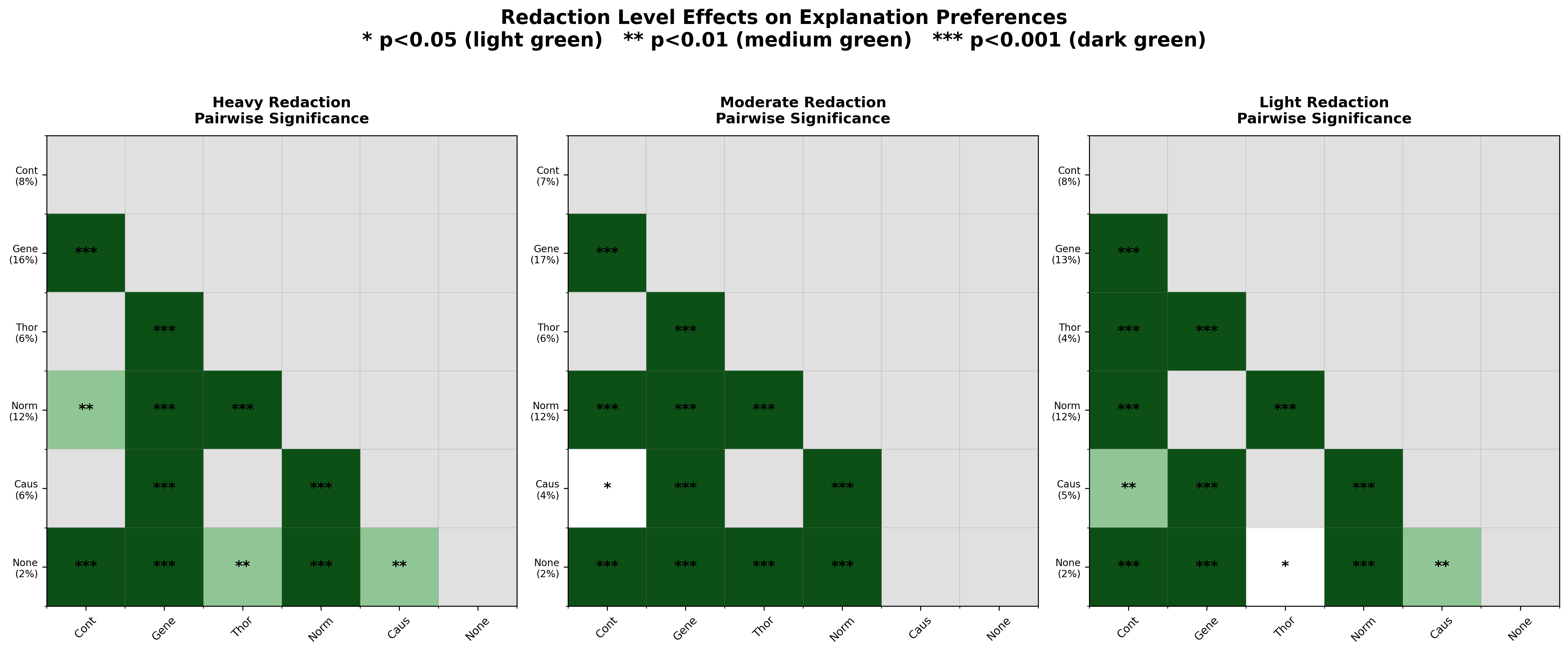}
    \caption{Pairwise comparison of explanation styles chosen for each redaction level. Significant differences are marked in light green, medium green, and dark green for $p<0.05$, $p<0.01$, and $p<0.001$, respectively.}
    \label{fig:rq1_redaction_analysis}
\end{figure*}

Figure \ref{fig:rq1_domain_analysis} shows the pairwise comparison of explanation styles chosen for each domain with significant differences marked. Figure \ref{fig:rq1_redaction_analysis} shows the pairwise comparison of explanation styles chosen for each redaction level, with significant differences marked.

\subsection{Interaction Effect between Domain and Redaction Level on Explanation Preferences}
\begin{figure*}[ht]
    \centering
    \includegraphics[width=\linewidth]{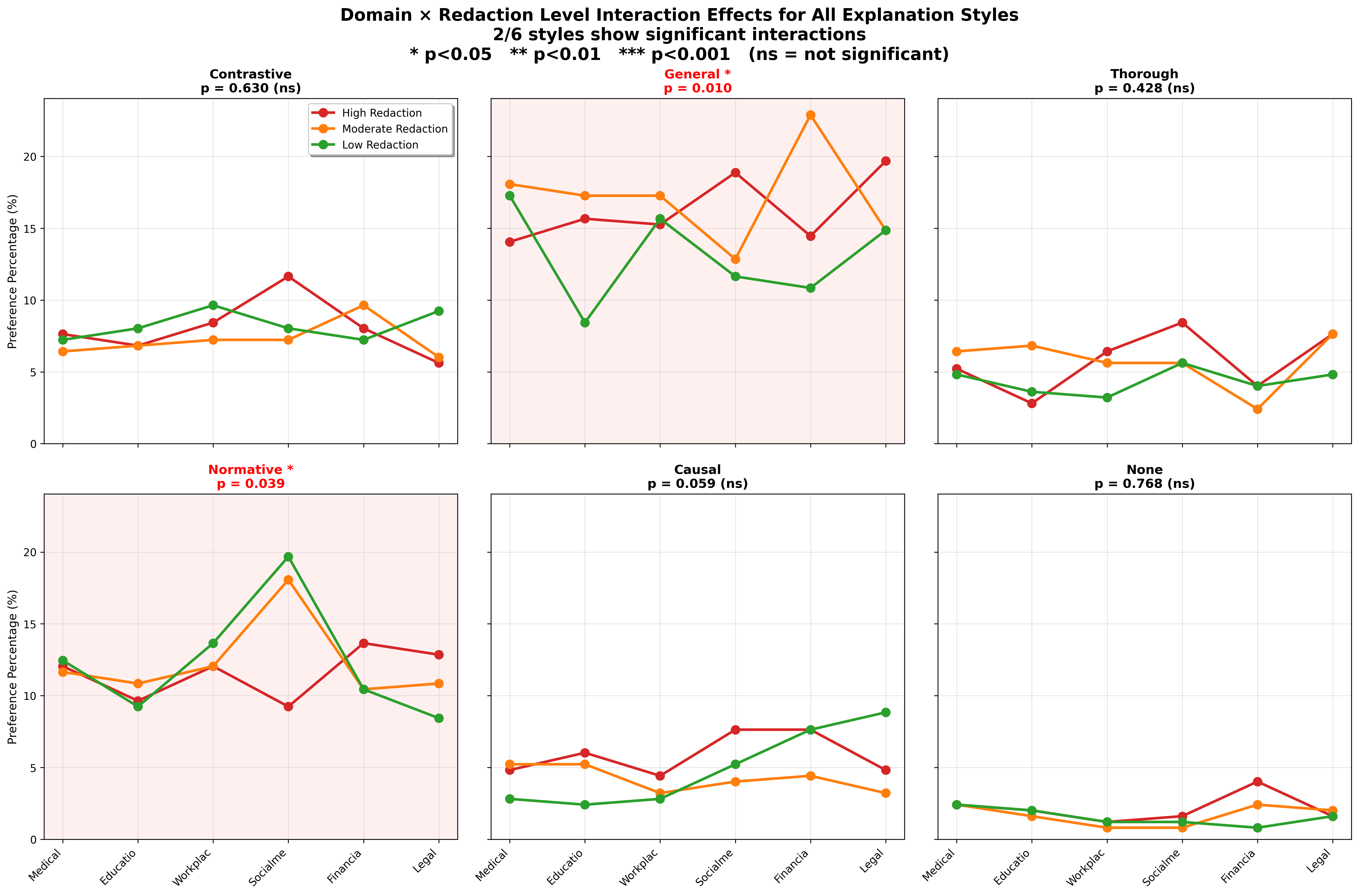}
    \caption{Domain and redaction level interaction for all explanation styles. Each subplot shows how the choice percentage varies for a particular style.}
    \label{fig:rq1_interaction_analysis}
\end{figure*}

Figure \ref{fig:rq1_interaction_analysis} shows how domain and redaction level affected participants' explanation preferences.

\end{document}